\documentclass[twocolumn]{aastex631}
\usepackage[utf8]{inputenc}
\usepackage{natbib}
\setcitestyle{round}
\usepackage{amsmath,amssymb}
\usepackage{afterpage}
\usepackage{xcolor}
\usepackage[caption=false]{subfig}
\usepackage{soul}
\usepackage{enumitem}
\usepackage{xcolor}

\newcommand{\hi}{{\rm H\,{\small I}}}
\newcommand{\hii}{{\rm H\,{\small II}}}
\newcommand{\kms}{\ensuremath{\,{\rm km\,s^{-1}}}}

\newcommand{\percc}{\ensuremath{\,{\rm cm^{-3}}}}

\newcommand{\persc}{\ensuremath{\,{\rm cm^{-2}}}}
\newcommand{\hcop}{\text{HCO\textsuperscript{+}}}

\newcommand{\hcn}{\text{HCN}}

\newcommand{\sio}{\text{SiO}}
\newcommand{\cch}{\text{C\textsubscript{2}H}}
\newcommand{\htwo}{\text{H\textsubscript{2}}}
\newcommand{\texc}{\ensuremath{T_{\rm{ex}}}}

\begin{document}

\title{Shaken or stirred: the diffuse interstellar medium with exceptionally high SiO abundance}

\correspondingauthor{Daniel R. Rybarczyk}
\email{rybarczyk@astro.wisc.edu}

\author[0000-0003-3351-6831]{Daniel R. Rybarczyk}
\affiliation{University of Wisconsin--Madison, Department of Astronomy, 475 N Charter St, Madison, WI 53703, USA}

\author[0000-0002-3418-7817]{Sne\v zana Stanimirovi\'c}
\affiliation{University of Wisconsin--Madison, Department of Astronomy, 475 N Charter St, Madison, WI 53703, USA}

\author[0000-0002-0354-1684]{Antoine Gusdorf}
\affiliation{Laboratoire de Physique de l'\'Ecole Normale Sup\'erieure, ENS, Universit\'e PSL, CNRS, Sorbonne Universit\'e, Universit\'e de Paris, 75005 Paris, France}
 \affiliation{Observatoire de Paris, PSL University, Sorbonne Universit\'e, LERMA, 75014 Paris, France}

\begin{abstract}
Interstellar shocks, a key element of stellar feedback processes, shape the structure of the interstellar medium (ISM) and are essential for the chemistry, thermodynamics, and kinematics of interstellar gas. Powerful, high-velocity shocks are driven by stellar winds, young supernova explosions, more evolved supernova remnants, cloud-cloud collisions, and protostellar outflows, whereas the existence and origin of much-lower-velocity shocks ($\lesssim10~\kms{}$) are not understood. Direct observational evidence for interstellar shocks in diffuse and translucent ISM environments have been especially lacking. We present the most sensitive survey to date of SiO---often considered an unambiguous tracer of interstellar shocks---in absorption, obtained with the Northern Extended Millimeter Array interferometer. We detect SiO in 5/8 directions probing diffuse and translucent environments without ongoing star formation. Our results demonstrate that SiO formation in the diffuse ISM (i.e., in the absence of significant star formation and stellar feedback) is more widespread and effective than previously reported. The observed SiO linewidths are all  $\lesssim4~\kms{}$, excluding high-velocity shocks as a formation mechanism. Yet, the SiO abundances we detect are mostly 1 to 2 orders of magnitude higher than typically assumed in quiescent environments and are often accompanied with other molecular transitions whose column densities cannot be explained with UV-dominated chemical models. Our results challenge the traditional view of SiO production via stellar-feedback sources and
emphasize the need for observational constraints on the distribution of Si in the gas phase and grain mantles, which are crucial for understanding the physics of grain processing and diffuse interstellar chemistry.
\end{abstract}

\section{Introduction} \label{sec:intro}
Shocks generated by various forms of stellar feedback play a pivotal role in galaxy evolution \citep{McKeeOstriker1977,Hopkins2012}. Spectral line observations of the SiO molecule have provided a key tool to identify possible feedback sources based on the abundances of SiO and shock velocities.  Shocks can enhance the SiO abundance, $X_{\rm{SiO}}=N(\mathrm{SiO})/ N(\htwo{})$, by up to six orders of magnitude relative to quiescent environments \citep{MartinPintado1992,AmoBaladron2011,LopezSepulcre2016,NaiPing2018,Li2019,ArmijosAbendano2020}.
High-velocity ($v\gtrsim25~\kms{}$) shocks caused by supernova remnants, large-scale cloud collisions, expanding \hii{} regions, or protostellar outflows are known to be highly effective at forming SiO \citep{MartinPintado1997,Gueth1998,Hatchell2001,Gusdorf2008} by extracting Si from the dust grains, which then reacts with either O$_2$ or OH to form SiO \citep{Herbst1989,Schilke1997}.
Recent observations, though, have provided surprising evidence for the existence and effectiveness of low velocity shocks ($v\lesssim15~\kms{}$) in forming SiO in dense interstellar environments \citep{JimenezSerra2010,NguyenLuong2013,DuarteCabral2014,Louvet2016,LopezSepulcre2016,DeSimone2022}.
The origin of these low velocity shocks remains poorly constrained, although possible explanations include converging flows \citep{JimenezSerra2010}, cloud collisions \citep{Louvet2016}, decelerated protostellar shocks from no-longer-active star formation events \citep{LopezSepulcre2016}, or trains of shocks formed at the boundary of an expanding bubble \citep{DeSimone2022}.

Observational measurements of SiO abundances are also important for constraining details of the dust grain physics. While it is understood that most of the Si in the ISM is locked inside the dust grain cores \citep{Jenkins2009}, the amount of Si that is either free in the gas-phase ISM or is located in the mantles of dust grains is currently poorly constrained, yet affects greatly many astrochemical reactions.
For example, modeling suggests that if $1$--$10\%$ of silicon is present in the gas phase, then SiO column densities produced in low-velocity shocks can be several orders of magnitude higher than environments where Si is entirely locked in grain cores and mantles \citep{NguyenLuong2013,Louvet2016}.

Shocks produced by stellar feedback predominantly shape dense environments most of which are star-forming. Meanwhile, in more diffuse environments without ongoing star formation to provide such feedback, chemical abundances have been puzzling for many years.
Standard astrochemical models have not been able to explain the high abundances of many species (\hcop{}, \hcn{}, and \cch{}, among others) in the diffuse ISM, leading to the hypothesis that turbulence, and in the most extreme cases shocks, could be responsible for the high abundances
\citep[e.g.,][]{Godard2009,Godard2010,ReachHeiles2021,Ryb2_2022}.
Yet, direct observations of shock tracers like SiO in the diffuse ($A_V\lesssim1$) and translucent ($1\lesssim A_V\lesssim4$) ISM have been extremely rare \citep{SiO2000,Corby2018}.

In this work, we search for \sio{} in absorption against 8 background radio continuum sources at $3~\rm{mm}$ wavelength in the direction of diffuse and translucent gas in the Milky Way with the Northern Extended Millimeter Array (NOEMA). We also observe \hcop{} in absorption in these directions to trace the total quantity of molecular gas. The observations are described in Section \ref{sec:observations}. In Section \ref{sec:methods} we present the methods for decomposing absorption spectra into Gaussian components and calculating the column densities of different molecular species. We then characterize the observed environments using new and existing data in the direction of our background radio continuum sources. In Section \ref{sec:results}, we report the \sio{} column densities and abundances in these diffuse and translucent environments. We discuss these results in Section \ref{sec:discussion}, both in the context of diffuse interstellar chemistry and in the context of the existing literature. We then present our conclusions in Section \ref{sec:conclusions}.

\section{Observations} \label{sec:observations}
\begin{figure*}
    \centering
    \includegraphics[width=\linewidth]{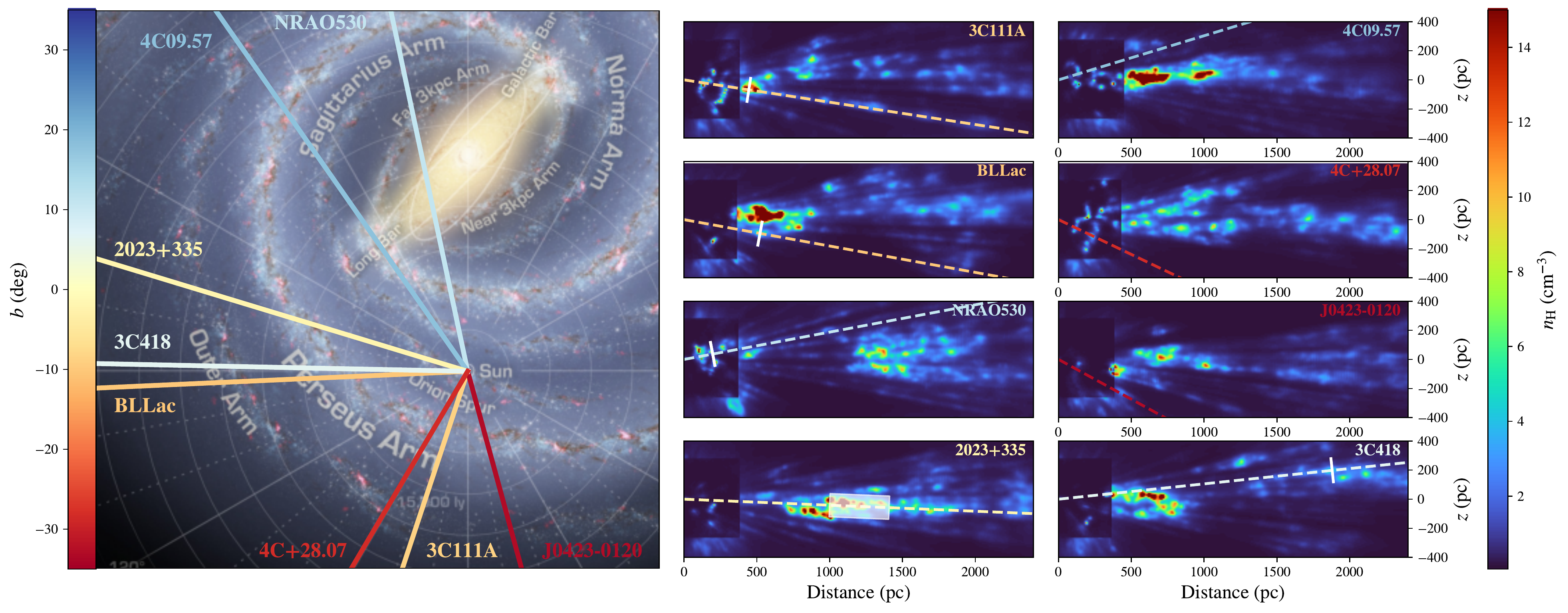}
    \caption{An overview of the sightlines observed in this work. \textit{Left:} artist's conception of the Milky Way Galaxy, with prominent features labeled (credit: NASA/JPL-Caltech/R. Hurt, SSC-Caltech). The directions of our extragalactic background sources are shown, colored according to Galactic latitude. \textit{Right:} cross-sections through the \citet{Lallement2019} and \citet{Leike2020} 3D dust maps of the local ISM. Both maps are shown with the same density scale; the peak density shown is $15~\percc{}$, but peak densities in the maps reach as high as $63~\percc{}$. The higher resolution \citet{Leike2020} map, visible at the shortest distances, is plotted over the lower-resolution \citet{Lallement2019} map. Due to the varying resolutions and methods for measuring $n_{\rm{H}}$, the densities may be discrepant between the two maps. The directions to our background sources are shown in dashed lines, colored by latitude, as in the left figure. The peak density in the direction of our background sources is $25~\percc{}$. White lines are drawn perpendicular to the sightlines to indicate at the estimated positions where SiO absorption is detected. For 2023+335, we show a partially-transparent box to indicate the rather large range of distances over which the SiO could presumably originate.}
    \label{fig:Sources}
\end{figure*}

We searched for \sio{} ($J=2-1$, 86.8470 GHz) and \hcop{} ($J=1-0$, 89.1885 GHz) in absorption with NOEMA (projects W19AQ, S20AB, and E20AA) in the direction of 8 background radio continuum sources, probing diffuse and translucent environments (Table \ref{tab:sources}; Figure \ref{fig:Sources}).
Sources were observed for a total of 1.1 to 11.8 hours (including overheads, spanning 1 to 5 sets of observations) to reach a targeted optical depth sensitivity of $\approx0.002$ (see Table \ref{tab:sources} for the achieved sensitivities). 
The bright quasars 3C84, 3C454.3, 3C345, 0059+581, 3C723, and 3C279 were used as bandpass calibrators; the bandpass was stable over the 0.9 to 5.1 hours observed in each set of observations.
Both lines were placed in high-resolution chunks ($62.5~\rm{kHz}$ channel spacing $\approx0.2~\kms{}$ velocity spacing) in the lower side band to achieve the optimal sensitivity. We carried out standard calibration using the CLIC and MAPPING software, part of the GILDAS software collection\footnote{https://www.iram.fr/IRAMFR/GILDAS} \citep{2005sf2a.conf..721P,2013ascl.soft05010G}. Because our background sources were bright ($\gtrsim1~\rm{Jy}$; Table \ref{tab:sources}), we further performed self-calibration using the 2020 Self-Calibration tool in MAPPING on all sources. The final spectra presented in Figure \ref{fig:spectra} were smoothed to 0.4 \kms{} velocity resolution. Table \ref{tab:sources} lists the noise in the line divided by continuum ($e^{-\tau(v)}$) for reference (the noise levels in the \hcop{} and \sio{} spectra are very similar). In the direction of 3C111, where the \hcop{} absorption is saturated, we also utilize observations of H$^{13}$CO$^+$ ($J=1-0$, 86.7543 GHz) from NOEMA obtained in the same observing run as the \hcop{} observations (project W19AQ), achieving comparable sensitivity.

\section{SiO and \hcop{} in the diffuse ISM} \label{sec:methods}
We detect \hcop{} in absorption in all 8 directions and we detect SiO in absorption in 5/8 directions in our sample.
In Figure \ref{fig:spectra} we show the observed SiO and \hcop{} absorption spectra with Gaussian fits; Table \ref{tab:gaussian_fits} lists all 31 \hcop{} Gaussian absorption components and all 8 \sio{} Gaussian absorption components (see Section \ref{subsec:gaussian_fits}).
Three of the absorbing structures are located within 500~pc, while the remaining 5 are at a distance of 1--2~kpc, as illustrated in Figure \ref{fig:Sources} (see Section \ref{sec:environments}). Because \hcop{} absorption in the direction of 3C111A is saturated, we also utilize H$^{13}$CO$^+$ observations from NOEMA in this direction \citep{LucasLiszt1998}; the H$^{13}$CO$^+$ spectrum is shown in Figure \ref{fig:spectra} and the Gaussian fits are listed in Table \ref{tab:gaussian_fits}.

\begin{table}
    \centering
    \begin{tabular}{|c|cc|c|c|c|}
    \hline
        Source & $\ell$ & $b$ & $F_{90}$ & $\sigma_{l/c}$ & $A_V$ \\
        & deg & deg & Jy & & mag \\
        \hline
        2023+335 & 73.1 & -2.4 & 1.01 & 0.001 & 5.21 \\
        3C418 & 88.8 & 6.0 & 1.00 & 0.001 & 3.18 \\
        3C111A & 161.7 & -8.8 & 1.14 & 0.002 & 4.41$^\dagger$ \\
        BL Lac & 92.6 & -10.4 & 2.10 & 0.001 & 1.22 \\
        NRAO 530 & 12.0 & 10.8 & 1.97 & 0.003 & 1.52 \\
        4C09.57 & 34.9 & 17.6 & 2.24 & 0.003 & 0.47 \\
        4C+28.07 & 149.5 & -28.5 & 0.97 & 0.002 & 0.45 \\
        J0423-0120 & 195.3 & -33.1 & 3.69 & 0.001 & 0.42 \\ 
        \hline
    \end{tabular}
    \caption{The background radio continuum sources against which we measured absorption in this work. We list the name of the continuum source (column 1), the Galactic longitude, $\ell$ (column 2), the Galactic latitude, $b$ (column 3), the 90 GHz flux density of the background source, $F_{90}$ (column 4), the noise in the line divided by the continuum ($\sim \sigma_{\tau}$; column 5), and the visual extinction along the line of sight, $A_V$ (column 6). All $A_V$ values are derived from \citet{Planck2016_GNILC} except 3C111A, indicated with a $\dagger$, which is derived from the higher-resolution map of California made with \textit{Herschel} by \citet{Lada2017}.}
    \label{tab:sources}
\end{table}

\begin{figure*}
\includegraphics[width=3.5in]{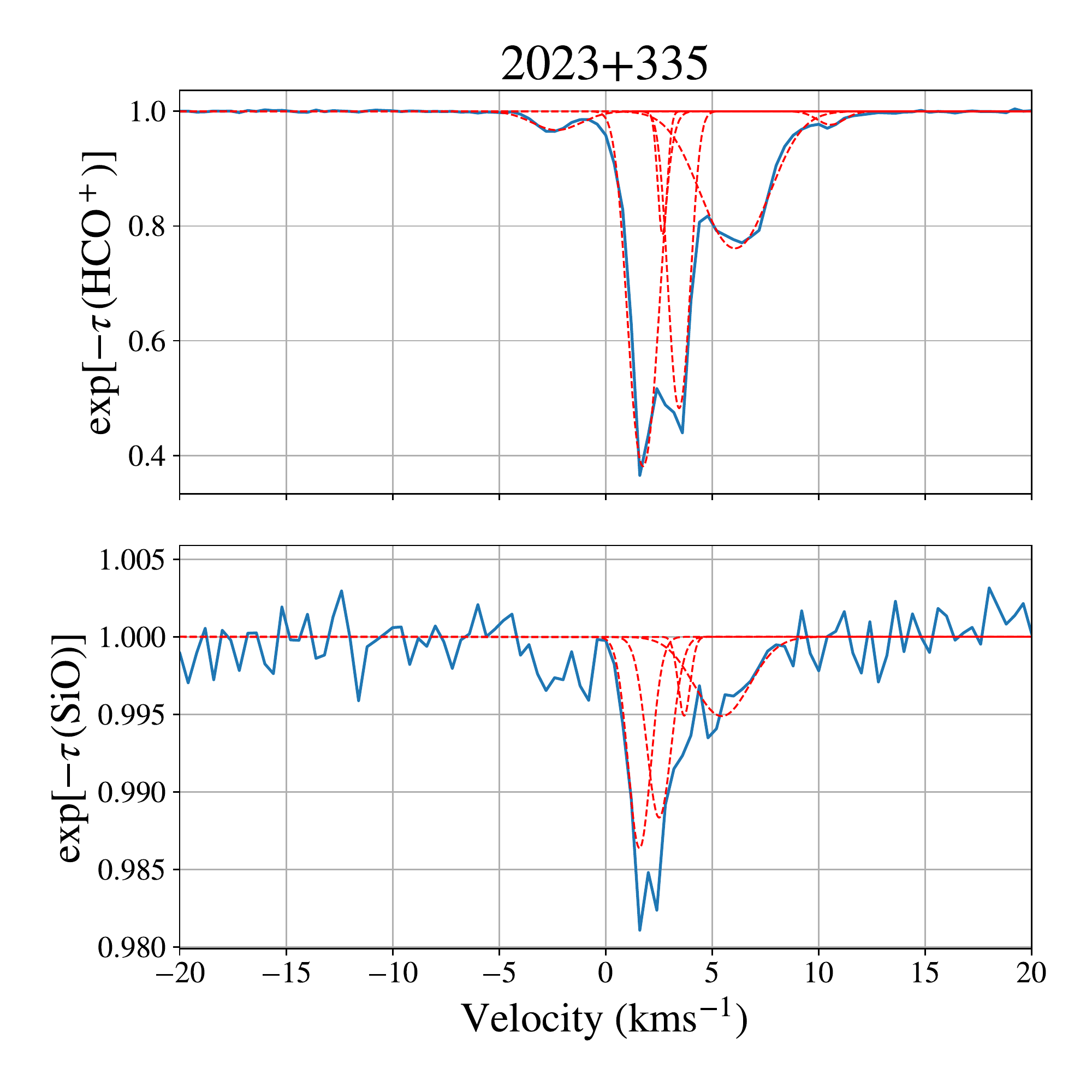}
\includegraphics[width=3.5in]{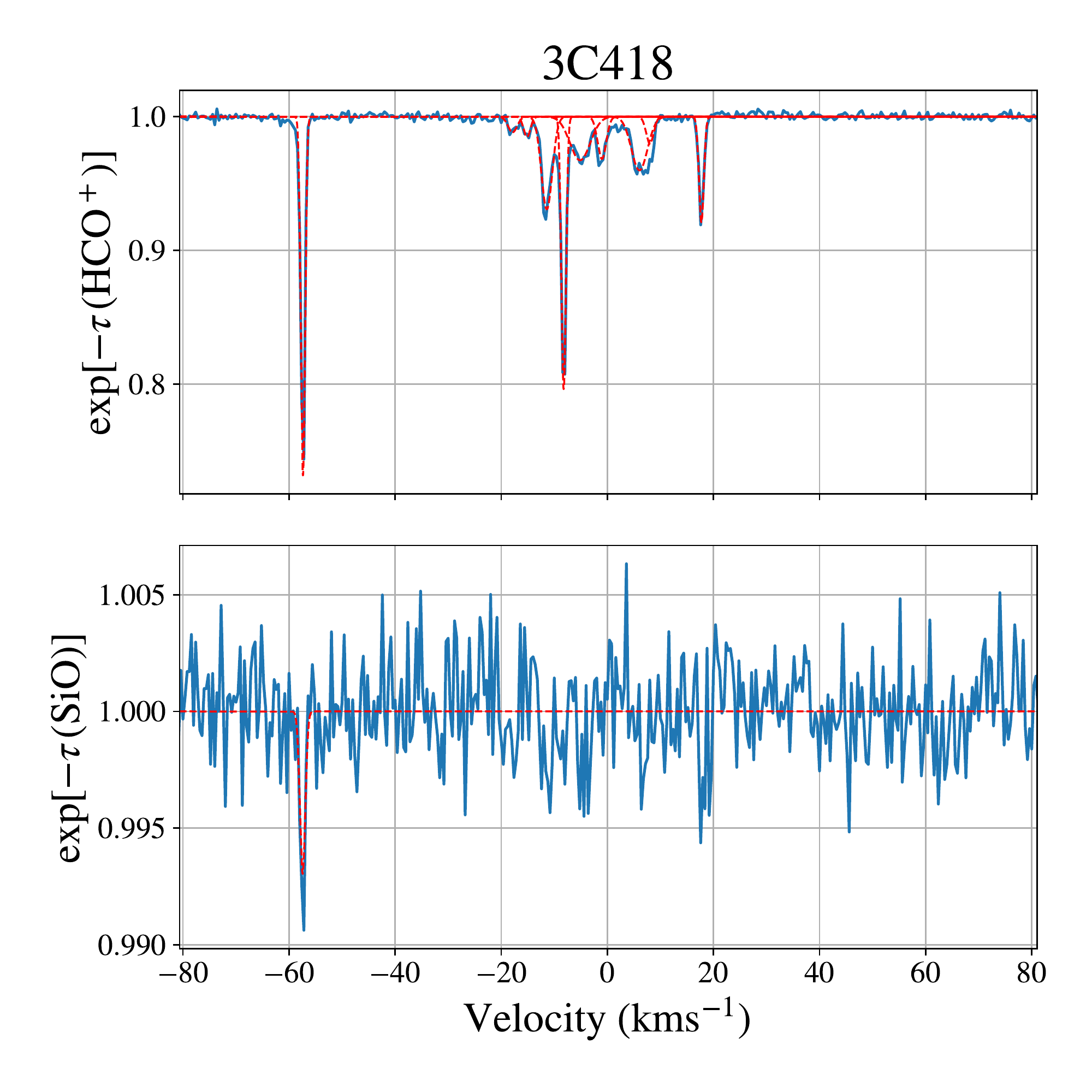}
\includegraphics[width=3.5in]{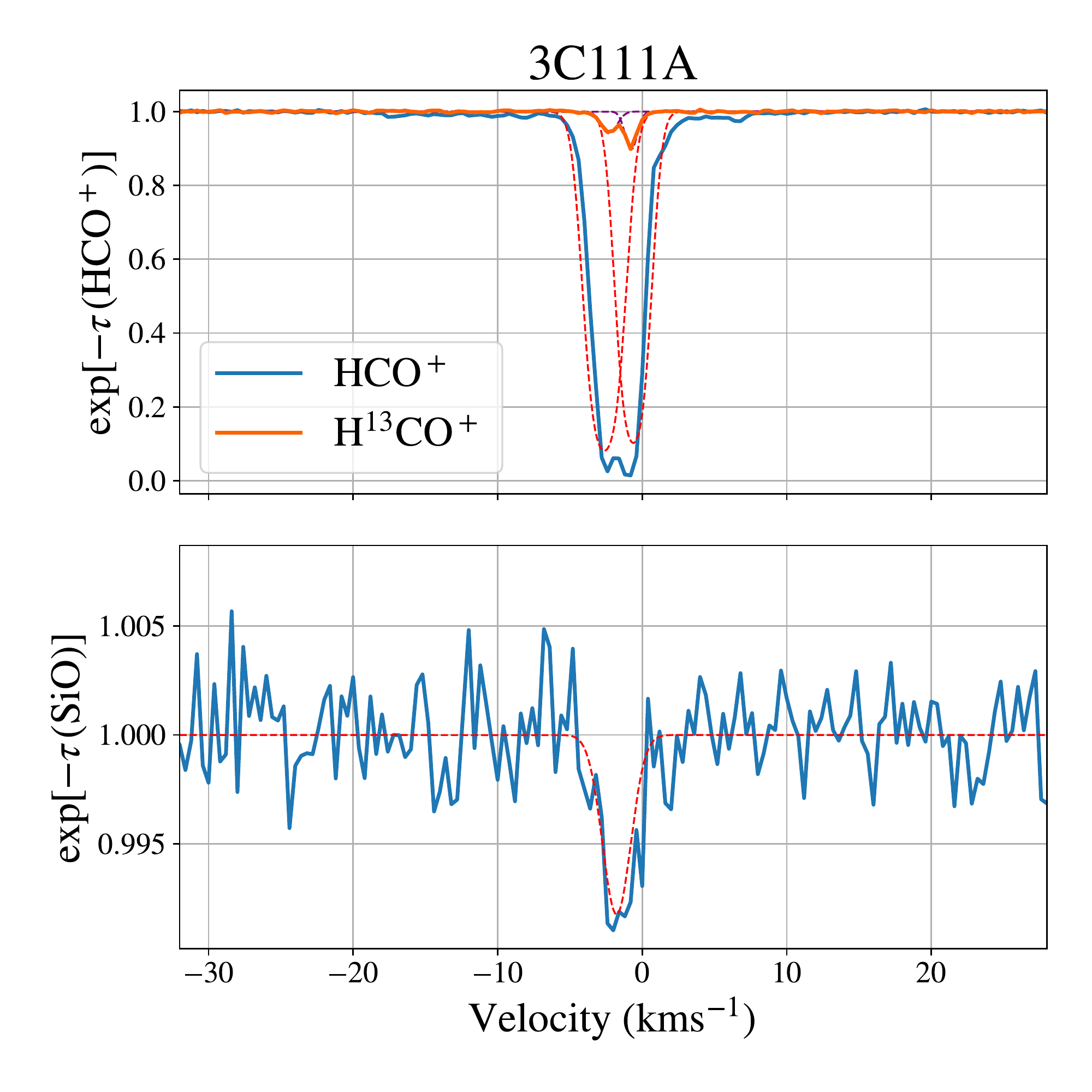}
\includegraphics[width=3.5in]{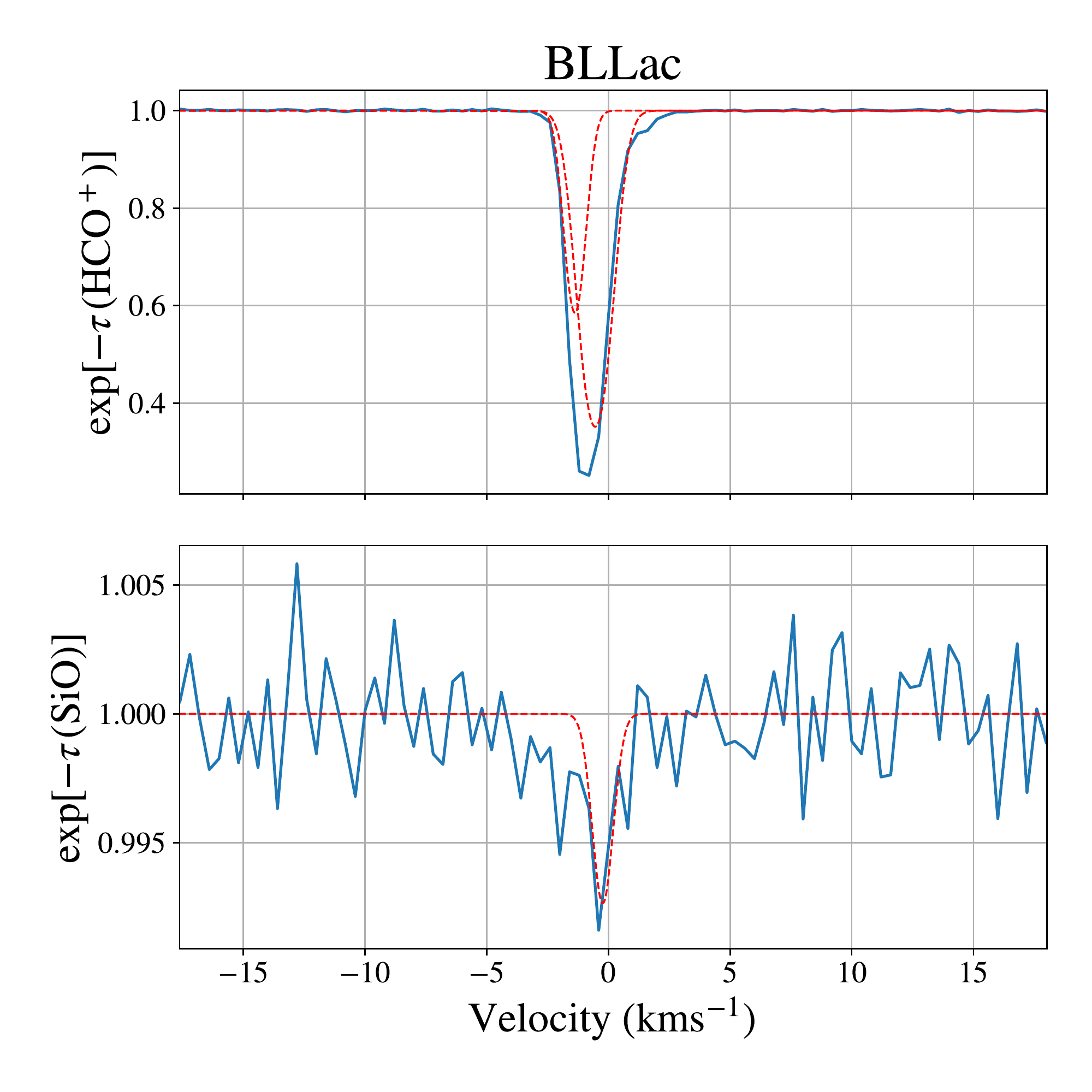}
\caption{}
\end{figure*}
\begin{figure*}\ContinuedFloat
\includegraphics[width=3.5in]{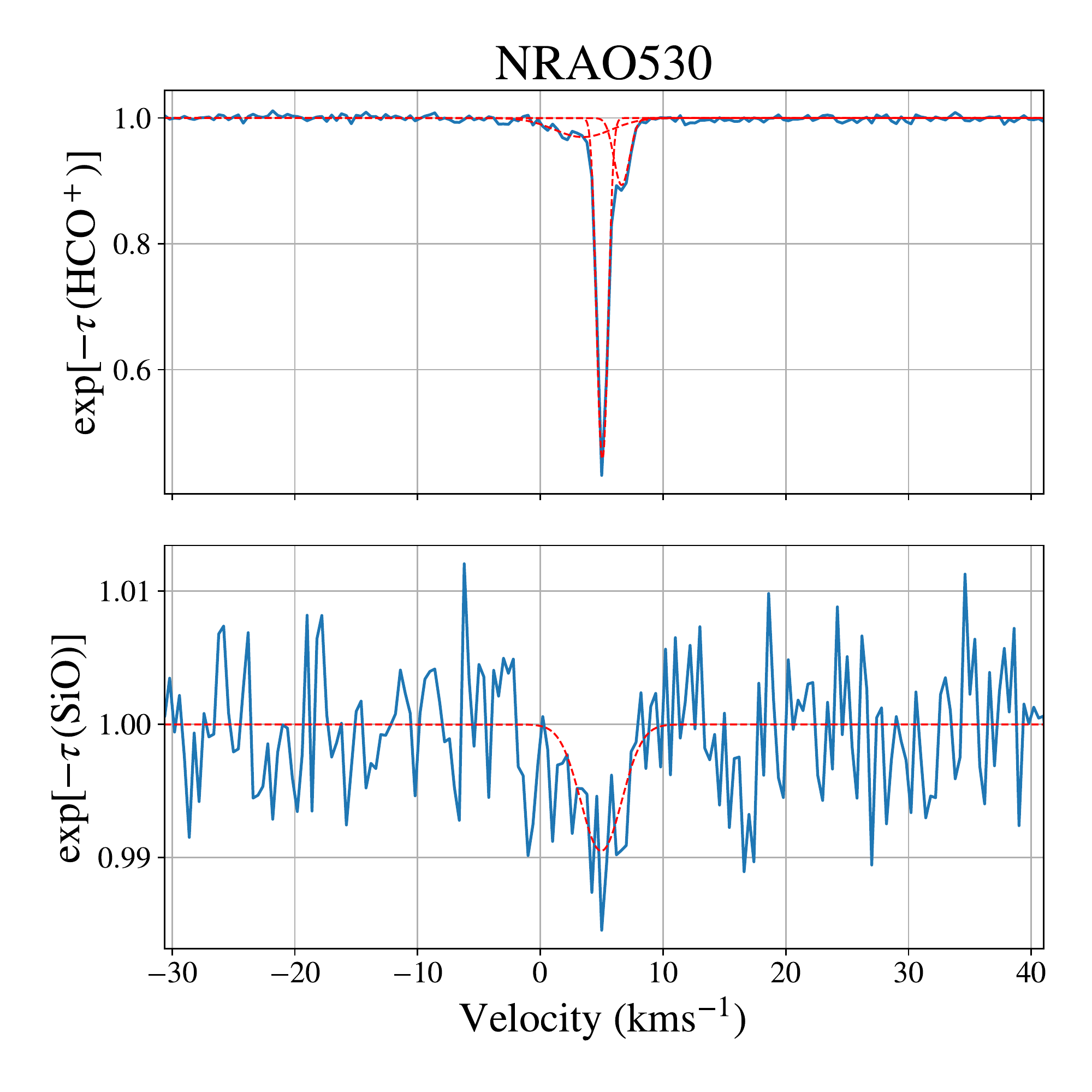}
\includegraphics[width=3.5in]{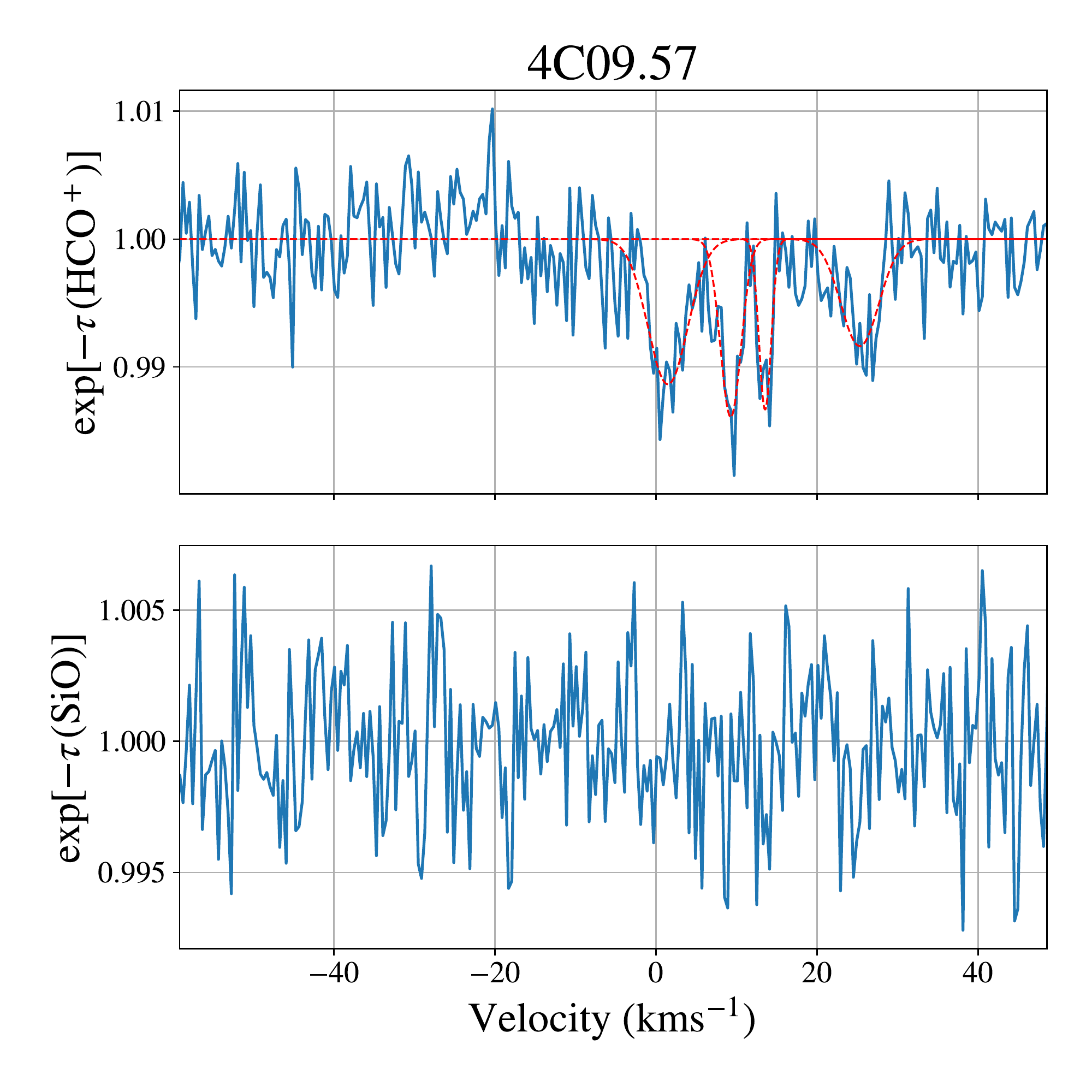}
\includegraphics[width=3.5in]{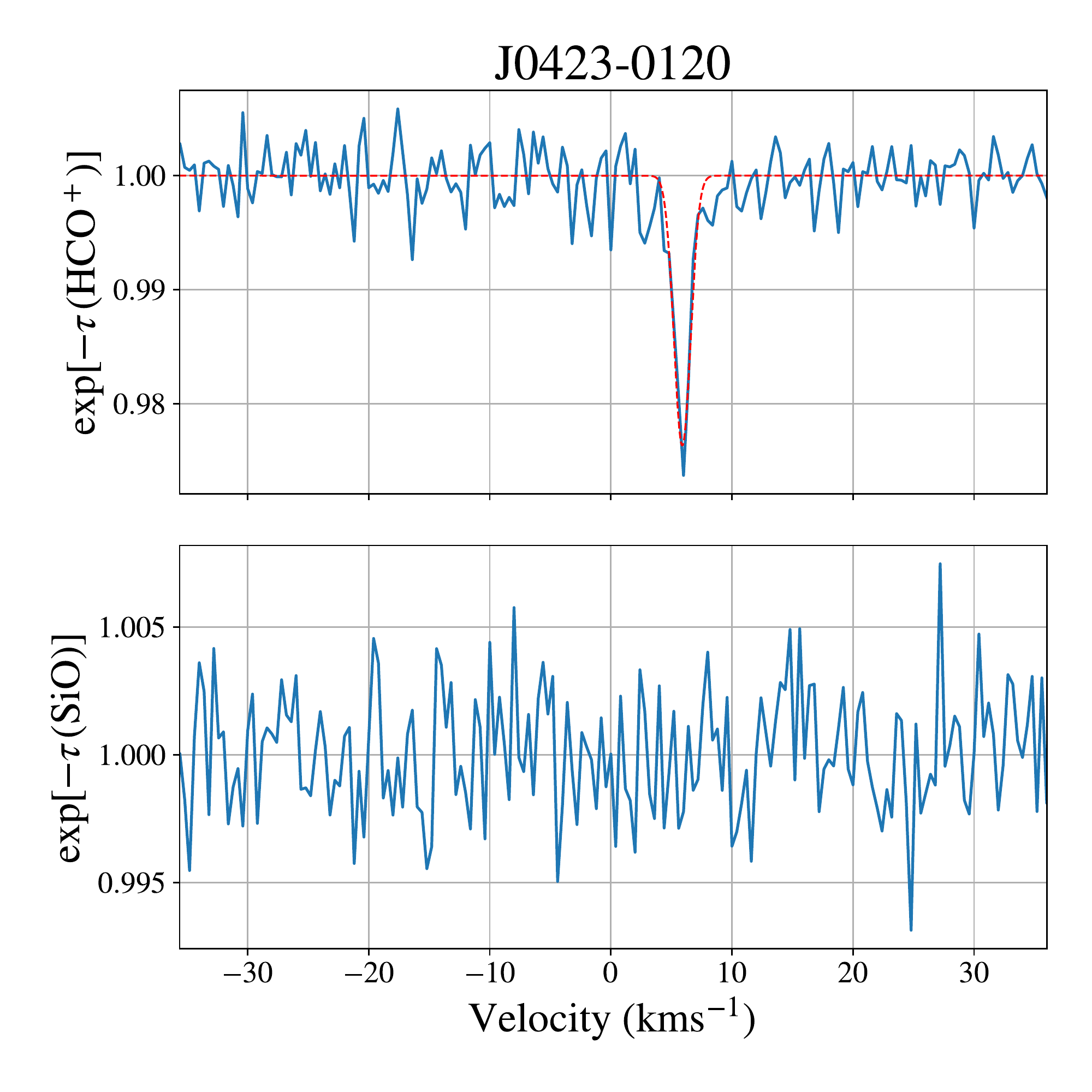}
\includegraphics[width=3.5in]{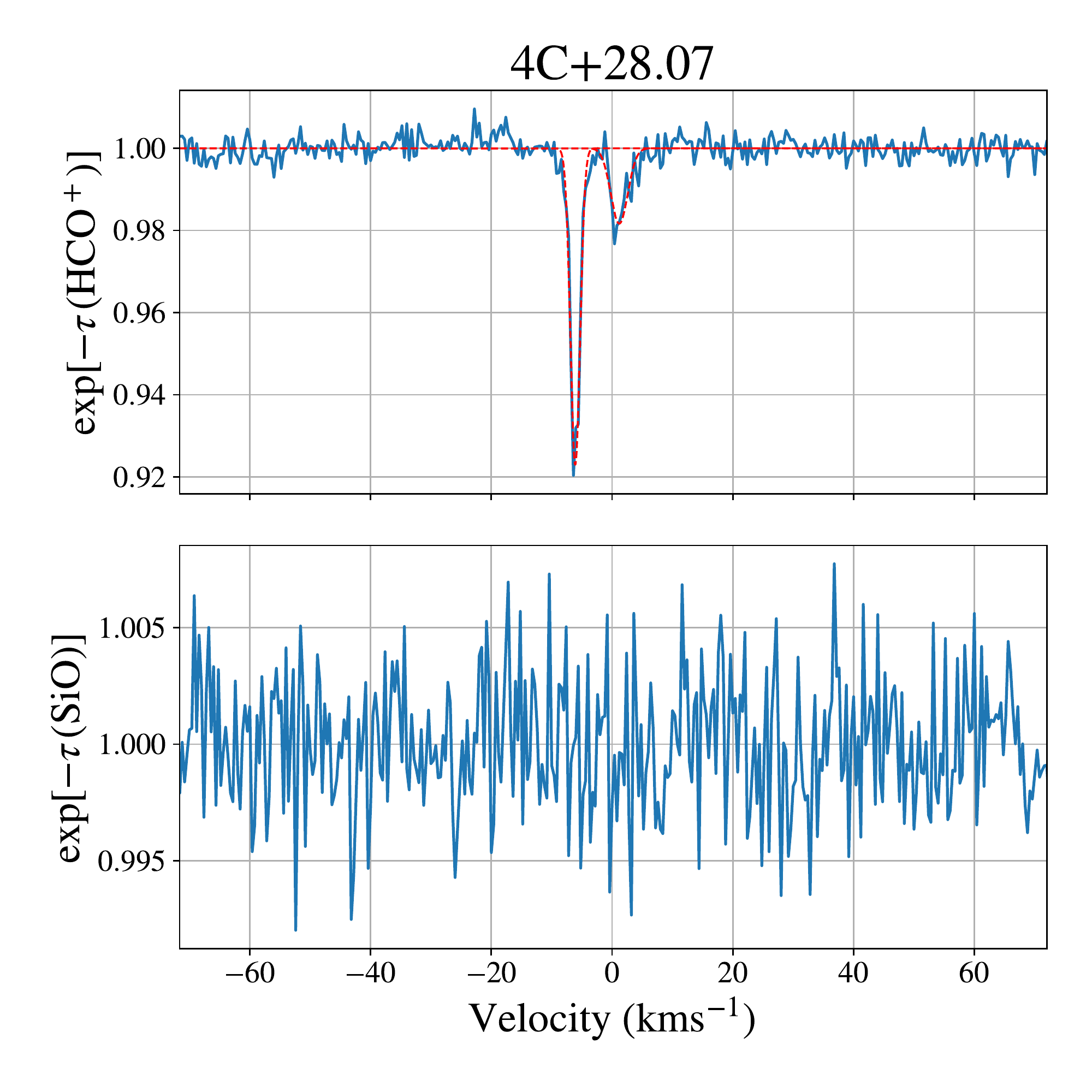}
\caption{The \hcop{} (\textit{top}) and \sio{} (\textit{bottom}) absorption spectra ($e^{-\tau}$) in the direction of our eight background sources. Gaussian fits are shown as red dashed lines for both species. For 3C111A, we also show the H$^{13}$CO$^+$ absorption spectra and Gaussian fits.}
\label{fig:spectra}
\end{figure*}

\subsection{Gaussian decomposition} \label{subsec:gaussian_fits}

We decompose all absorption spectra in this work into Gaussian components. For each spectrum, we supplied initial guesses for the number of components as well as the amplitude, full width at half-maximum (FWHM), and central velocity of each component. We then used Python's \texttt{scipy.optimize.curve\_fit} to fit the Gaussian components to the spectrum. 
To obtain the final estimates and uncertainties, we re-fit the spectra with Gaussian components over 1000 trials, adding random noise to the spectrum in each trial. We took the mean and standard deviation over all trials to be the estimate and uncertainty for each fitted quantity (the amplitude, FWHM, and central velocity of each Gaussian component). 

We used the same initial guesses for the number of components and the central velocities of components in the \hcop{} and \sio{} absorption spectra. The number of components was determined by inspection of the \hcop{} absorption spectrum for each sightline, which assumes that any structures observed in \sio{} are also observed in \hcop{} \citep[visual inspection  clearly shows that all peaks in \sio{} absorption are coincident with peaks in \hcop{} absorption; moreover, previous observations confirm \hcop{} is present where \sio{} is observed;][]{SiO2000,LopezSepulcre2016}. We then rejected components detected at a level $<3\sigma$.
The only exception is in the case of NRAO 530, where the peak in the \sio{} absorption is detected only at a level of $2.2\sigma$, while the integrated absorption ($\int \tau(v)\,dv$) is detected at $2.7\sigma$. Both metrics suggest that this is only a marginal detection. Yet, because the peak in the \sio{} absorption is coincident with the peak in the \hcop{} absorption and because previous work has revealed excess \sio{} absorption toward NRAO 530 at $\sim5~\kms{}$ \citep[see Section \ref{subsec:previous_obs};][]{SiO2000}, we classify this feature as a detection.
All of the Gaussian fits are shown in Figure \ref{fig:spectra} and listed in Table \ref{tab:gaussian_fits}. In Section \ref{subsubsec:XSiO}, we discuss the bias to $X_{\sio{}}$ introduced by our Gaussian fitting. 

The \hcop{} absorption in the direction of 3C111A is saturated, so the approach outlined above is potentially problematic. We therefore also fit Gaussian components to the (unsaturated) H$^{13}$CO$^+$ absorption spectrum. The H$^{13}$CO$^+$ absorption spectrum is shown in Figure \ref{fig:spectra} and the Gaussian components are listed in \ref{tab:gaussian_fits} (indicated with a $\dagger$). We estimate $N(\hcop{})=61\times N(\rm{H^{13}CO^+})$ based on previous observations of \hcop{} and \hcn{} istopologues in the direction of 3C111 \citep{LucasLiszt1998}. The fitted H$^{13}$CO$^+$ absorption components have narrower linewidths than their saturated \hcop{} counterparts, but the column densities vary by $\lesssim30\%$ when using H$^{13}$CO$^+$ compared to using just \hcop{}, so do not change any of the qualitative results presented in this work. The results presented in Section \ref{sec:results} are those derived from H$^{13}$CO$^+$, which we take to be more reliable. The uncertainties include the uncertainty in $N(\hcop{})/ N(\rm{H^{13}CO^+})$.

\subsection{Deriving molecular column densities}\label{subsec:molecular_column_densities}
For a transition from an upper state $u$ to a lower state $l$ observed in absorption, the column density, $N$, is given by
\begin{equation} \label{eq:N}
    N = Q(\texc{})\frac{8\pi\nu^3}{c^3}\frac{1}{g_u A_{ul}}\Big[1-\exp(-h\nu/k\texc{})\Big]^{-1} \int\tau(v) dv,
\end{equation}
where \texc{} is the excitation temperature (see Section \ref{subsubsec:Tex}), $Q(\texc{})$ is the partition function, $\nu$ is the rest frequency of the transition, $g_u$ is the degeneracy of the upper state, $A_{ul}$ is the Einstein $A$ coefficient for the transition, and $\tau(v)$ is the optical depth of the line at velocity $v$. We evaluate Equation \ref{eq:N} using values given in the Cologne Database for Molecular Spectroscopy \citep[CDMS;][]{2001A&A...370L..49M,2016JMoSp.327...95E} and the Leiden Atomic and Molecular Database \citep[LAMDA;][]{2010ascl.soft10077S}. Results are listed in Table \ref{tab:N_tau_relations} for different excitation temperatures (in this work, we use $\texc{}=2.725~\rm{K}$; see next section).

\begin{deluxetable*}{|c|ccc|ccc|}
\tablehead{
\colhead{Sightline} & \colhead{$\tau_{\hcop{}}$} & \colhead{\textbf{$v_{\hcop{}}$}} & \colhead{$\Delta v_{\hcop{}}$} & \colhead{$\tau_{\sio{}}$} & \colhead{\textbf{$v_{\sio{}}$}} & \colhead{$\Delta v_{\sio{}}$}}
\tablecaption{Gaussian components fitted to the \hcop{} (columns 2--4) and \sio{} (columns 5--7) absorption spectra. The optical depth ($\tau$), the FWHM ($\Delta v$), and the central velocity ($v$) of each component are listed for both molecular species. For 3C111A, we also list results for the Gaussian decomposition of the H$^{13}$CO$^+$ spectrum, indicated with a $\dagger$ superscript. \label{tab:gaussian_fits}}
\startdata
2023+335 & $0.965\pm0.001$ & $1.75\pm0.00$ & $1.40\pm0.00$& $0.014\pm0.001$ & $1.59\pm0.11$ & $1.25\pm0.17$ \\ 
& $0.243\pm0.003$ & $2.67\pm0.00$ & $0.50\pm0.00$& $0.012\pm0.002$ & $2.51\pm0.12$ & $1.35\pm0.34$ \\ 
& $0.729\pm0.001$ & $3.45\pm0.00$ & $1.05\pm0.00$& $0.005\pm0.001$ & $3.68\pm0.15$ & $0.67\pm0.19$ \\ 
& $0.273\pm0.001$ & $6.07\pm0.01$ & $3.59\pm0.01$& $0.005\pm0.001$ & $5.45\pm0.04$ & $3.07\pm0.74$ \\ 
& $0.024\pm0.001$ & $10.56\pm0.03$ & $1.50\pm0.06$& $\cdot\cdot\cdot$ & $\cdot\cdot\cdot$ & $\cdot\cdot\cdot$ \\ 
& $0.033\pm0.001$ & $-2.30\pm0.00$ & $2.66\pm0.00$& $\cdot\cdot\cdot$ & $\cdot\cdot\cdot$ & $\cdot\cdot\cdot$ \\ 
& $0.007\pm0.001$ & $-81.49\pm0.17$ & $2.63\pm0.72$& $\cdot\cdot\cdot$ & $\cdot\cdot\cdot$ & $\cdot\cdot\cdot$ \\ 
\hline
3C111A & $2.514\pm0.003$ & $-2.62\pm0.01$ & $2.18\pm0.01$& $0.008\pm0.002$ & $-1.77\pm0.16$ & $2.30\pm0.35$ \\ 
& $0.061\pm0.002^\dagger$ & $-2.30 \pm 0.02^\dagger$ & $1.29\pm 0.07^\dagger$ &   &   &   \\ 
& $2.280\pm0.004$ & $-0.62\pm0.01$ & $1.92\pm0.01$& $\cdot\cdot\cdot$ & $\cdot\cdot\cdot$ & $\cdot\cdot\cdot$ \\ 
& $ 0.108 \pm 0.002^\dagger$ & $ -0.76 \pm 0.01^\dagger $ & $ 0.97 \pm 0.03^\dagger $ & & & \\
\hline
3C418 & $0.313\pm0.001$ & $-57.38\pm0.00$ & $0.91\pm0.01$& $0.007\pm0.002$ & $-57.44\pm0.10$ & $1.20\pm0.19$ \\ 
& $0.012\pm0.001$ & $-17.86\pm0.12$ & $2.01\pm0.32$& $\cdot\cdot\cdot$ & $\cdot\cdot\cdot$ & $\cdot\cdot\cdot$ \\ 
& $0.015\pm0.001$ & $-15.14\pm0.08$ & $1.49\pm0.17$& $\cdot\cdot\cdot$ & $\cdot\cdot\cdot$ & $\cdot\cdot\cdot$ \\ 
& $0.072\pm0.001$ & $-11.41\pm0.04$ & $2.38\pm0.10$& $\cdot\cdot\cdot$ & $\cdot\cdot\cdot$ & $\cdot\cdot\cdot$ \\ 
& $0.228\pm0.004$ & $-8.23\pm0.00$ & $0.98\pm0.03$& $\cdot\cdot\cdot$ & $\cdot\cdot\cdot$ & $\cdot\cdot\cdot$ \\ 
& $0.033\pm0.002$ & $-5.14\pm0.18$ & $4.45\pm0.99$& $\cdot\cdot\cdot$ & $\cdot\cdot\cdot$ & $\cdot\cdot\cdot$ \\ 
& $0.032\pm0.002$ & $-1.02\pm0.04$ & $1.89\pm0.15$& $\cdot\cdot\cdot$ & $\cdot\cdot\cdot$ & $\cdot\cdot\cdot$ \\ 
& $0.041\pm0.001$ & $6.05\pm0.07$ & $3.54\pm0.15$& $\cdot\cdot\cdot$ & $\cdot\cdot\cdot$ & $\cdot\cdot\cdot$ \\ 
& $0.020\pm0.002$ & $8.03\pm0.06$ & $1.65\pm0.08$& $\cdot\cdot\cdot$ & $\cdot\cdot\cdot$ & $\cdot\cdot\cdot$ \\ 
& $0.082\pm0.001$ & $17.73\pm0.01$ & $1.06\pm0.02$& $\cdot\cdot\cdot$ & $\cdot\cdot\cdot$ & $\cdot\cdot\cdot$ \\ 
\hline
4C+28.07 & $0.080\pm0.002$ & $-6.10\pm0.01$ & $1.86\pm0.03$& $\cdot\cdot\cdot$ & $\cdot\cdot\cdot$ & $\cdot\cdot\cdot$ \\ 
& $0.019\pm0.002$ & $1.18\pm0.11$ & $3.30\pm0.33$& $\cdot\cdot\cdot$ & $\cdot\cdot\cdot$ & $\cdot\cdot\cdot$ \\ 
\hline
4C09.57 & $0.011\pm0.003$ & $1.47\pm0.27$ & $5.82\pm1.60$& $\cdot\cdot\cdot$ & $\cdot\cdot\cdot$ & $\cdot\cdot\cdot$ \\ 
& $0.014\pm0.003$ & $9.27\pm0.21$ & $3.01\pm0.73$& $\cdot\cdot\cdot$ & $\cdot\cdot\cdot$ & $\cdot\cdot\cdot$ \\ 
& $0.013\pm0.003$ & $13.54\pm0.18$ & $1.77\pm0.27$& $\cdot\cdot\cdot$ & $\cdot\cdot\cdot$ & $\cdot\cdot\cdot$ \\ 
& $0.008\pm0.003$ & $25.29\pm0.34$ & $5.54\pm1.55$& $\cdot\cdot\cdot$ & $\cdot\cdot\cdot$ & $\cdot\cdot\cdot$ \\ 
\hline
BLLac & $0.536\pm0.001$ & $-1.38\pm0.00$ & $0.96\pm0.01$& $\cdot\cdot\cdot$ & $\cdot\cdot\cdot$ & $\cdot\cdot\cdot$ \\ 
& $1.046\pm0.002$ & $-0.55\pm0.00$ & $1.45\pm0.00$& $0.007\pm0.002$ & $-0.23\pm0.16$ & $1.01\pm0.34$ \\ 
\hline
J0423-0120 & $0.024\pm0.001$ & $5.90\pm0.03$ & $1.62\pm0.10$& $\cdot\cdot\cdot$ & $\cdot\cdot\cdot$ & $\cdot\cdot\cdot$ \\ 
\hline
NRAO530 & $0.780\pm0.003$ & $5.05\pm0.00$ & $0.96\pm0.01$& $0.010\pm0.004$ & $4.96\pm0.33$ & $3.92\pm1.13$ \\ 
& $0.113\pm0.003$ & $6.63\pm0.02$ & $1.44\pm0.05$& $\cdot\cdot\cdot$ & $\cdot\cdot\cdot$ & $\cdot\cdot\cdot$ \\ 
& $0.031\pm0.003$ & $3.47\pm0.19$ & $5.44\pm0.45$& $\cdot\cdot\cdot$ & $\cdot\cdot\cdot$ & $\cdot\cdot\cdot$ \\ 
\enddata
\end{deluxetable*}
\vspace{-1cm}
\subsection{Excitation temperatures} \label{subsubsec:Tex}
According to Equation \ref{eq:N}, the excitation temperature is crucial for determining the column density from absorption measurements.
For most molecular species in the diffuse and translucent ISM, \texc{} is typically assumed to be equal to the temperature of the CMB, 2.725 K \citep{LucasLiszt1996,Liszt2001HCN}.
Indeed, for sightlines where both the $(2-1)$ and $(1-0)$ or the $(3-2)$ and $(1-0)$ transitions of \hcop{} were observed in absorption, the \hcop{} excitation temperature was found to be equal to $T_{\rm{CMB}}$ within errors \citep{Godard2010,Luo2020}. 

Nevertheless, numerical simulations have shown that the \hcop{} excitation temperature may rise above $T_{\rm{CMB}}$ to $\sim4\text{--}5~\rm{K}$ for densities $300~\percc{}\lesssim n \lesssim10^{3}~\percc{}$ \citep{Gong2020,Ryb2_2022}. Such behavior could lead to the \hcop{} column densities being underestimated by factor of $\sim2$.
Meanwhile, \sio{} excitation temperatures are usually thought to be higher than $T_{\rm{CMB}}$, with estimates---and in some cases direct measurements---ranging from roughly $5~\rm{K}\text{--}100~\rm{K}$, mostly in the range of $\sim10~\rm{K}\text{--}40~\rm{K}$ \citep{JimenezSerra2010,LopezSepulcre2016,NaiPing2018,Rivilla2018,Li2019,AA2020}, but these are in much higher-density environments ($10^4~\percc{}\lesssim n \lesssim10^7~\percc{}$) than probed by our observations.

In this work, we assume that the excitation temperature of \hcop{} is equal to $T_{\rm{CMB}}$, to be consistent with previous work \citep{LucasLiszt1996,Liszt2001HCN,Liszt2014,Ryb1_2022} and with existing observations of different rotational transitions of \hcop{} \citep{Godard2010,Luo2020}, as well as the weakness of the \hcop{} emission in these directions \citep[and Section \ref{subsec:densities}]{LucasLiszt1996}.
We also assume that the excitation temperature of SiO is equal to $T_{\rm{CMB}}$, as suggested by the RADEX non-LTE molecular radiative transfer code \citep{vanderTak2007} for the low densities in the directions of our background sources (see Section \ref{sec:environments}). Results for $\texc{}=4~\rm{K}$ and $\texc{}=5~\rm{K}$ are listed in Table \ref{tab:N_tau_relations} for reference.
\begin{table}
    \centering
    \begin{tabular}{|c|c|c|}
    \hline
        $T_{\rm{ex}}$ & $N/\int\tau_{\hcop{}} dv$ & $N/\int\tau_{\sio{}}dv$ \\
        $\rm{K}$ & \persc{}/\kms{} & \persc{}/\kms{} \\
        \hline
        $2.725$ & $1.09\times10^{12}$ & $3.45\times10^{12}$\\
        $4$ & $1.77\times10^{12}$ & $5.53\times10^{12}$ \\
        $5$ & $2.44\times10^{12}$ & $7.53\times10^{12}$ \\ 
        \hline
    \end{tabular}
    \caption{Conversions between the column density, $N$, and integrated optical depth, $\int\tau dv$, for \hcop{} and \sio{} at different excitation temperatures. In this work, we use $\texc{}=2.725~\rm{K}$ (see discussion in Section \ref{subsubsec:Tex}).}
    \label{tab:N_tau_relations}
\end{table}

\subsection{Measuring $N(\htwo{})$ and SiO abundances} \label{subsubsec:XSiO}
For every component identified in \sio{} absorption, we measure the \sio{} abundance, $X_{\sio{}}=N(\sio{})/N(\htwo{})$, where $N(\htwo{})\approx N(\hcop{})/3\times10^{-9}$ \citep[e.g.,][see discussion below]{LisztLucas2000,Liszt2010}. $N(\sio{})$ and $N(\hcop{})$ are solved using Equation \ref{eq:N} for the Gaussian components. There are multiple Gaussian \hcop{} absorption components identified along all lines of sight where \sio{} is detected in absorption. In order to isolate individual absorbing structures along the line of sight, we calculate $X_{\sio{}}$ using the \hcop{} absorption component closest in velocity to each \sio{} absorption component. 

The conversion from $N(\hcop{})$ to $N(\htwo{})$ is essential for calculating \sio{} abundances in this work. 
A large body of work using high-resolution interferometric observations exists and has confirmed a
linear relationship between the \hcop{} and \htwo{} column densities in the diffuse/translucent medium \citep{LisztLucas2000,Liszt2010}, with recent results placing the conversion factor at $N(\hcop{})/N(\htwo{})=3\times10^{-9}\pm 0.21~\rm{dex}$ \citep{Gerin2019,LisztGerin2023}. We adopt this empirically-derived result in this work. These estimates are derived from interferometric observations of \hcop{} in absorption, which are superior to single-dish studies because they probe very small solid angles and trace even cold, weakly excited molecular gas that does not emit significant radiation. As a confirmation that this linear relation is well constrained, it has been shown that the column density of \htwo{} inferred from \hcop{} (using a conversion factor of $3\times10^{-9}$) could account for the difference between the column densities derived from \hi{} and CO (another proxy for \htwo{}) and those inferred from $\gamma$ rays and dust emission \citep[considered to be unbiased tracers of the total column density][]{Liszt2018}. 

For all sightlines except 3C418, there are $\geq2$ \hcop{} components at similar velocities to where we see \sio{} absorption. This clearly introduces potential biases, as the calculation of $X_{\sio{}}$ depends on our Gaussian decomposition (of both the \sio{} and the \hcop{} spectra) as well as our velocity matching. 
In order to quantify this bias, we also estimate $X_{\sio{}}$ integrated over the total velocity range where \sio{} and \hcop{} are observed, rather than just for the individual Gaussian components. We summarize these results in Table \ref{tab:XSiO_no_Gauss}. For 3C111A (using both \hcop{} and H$^{13}$CO$^{+}$), BL Lac, and NRAO 530, $X_{\rm{SiO}}$ estimated over the entire velocity interval is within a factor of 2 of that estimated for the Gaussian components. There is a larger variation across 2023+335, although the Gaussian component with the highest-estimated $X_{\sio{}}$ accounts for only a small fraction of the total \sio{} and \hcop{} column densities along the line of sight, and the other three components are within a factor of 3 of $X_{\sio{}}$ evaluated over the entire velocity range. Thus, our choice of Gaussian fits may bias our overall estimate of the \sio{} abundances along the diffuse and translucent lines of sight in our sample by a typical factor of $<3$. This does not meaningfully alter our conclusion that the \sio{} abundances we measure are significantly higher than those typically quoted for quiescent environments (Section \ref{sec:results}).

\begin{table*}
    \centering
    \begin{tabular}{|c|c|c|c|}
    \hline
        Sightline & Velocity range & $X_{\sio{}}$ - Full velocity range & $X_{\sio{}}$ - Gaussian decomposition \\
        & \kms{} & $10^{-10}$ & $10^{-10}$ \\
        \hline
        2023+335  &  (-10, 20)  &  $1.6$ &  $0.54,1.6,1.1,10.5$\\
        3C111A  &  (-10, 10)  &  $0.18^\dagger$ & $0.33$ \\
        BL Lac  &  (-5, 5)  &  $0.75$ & $0.47$ \\
        NRAO 530  &  (-5, 10)  &  $4.5$ & $4.7$ \\
        \hline
    \end{tabular}
    \caption{For sightlines where we detect SiO in absorption (column 1), the SiO abundance, $X_{\sio{}}$, evaluated over the entire velocity range (column 2) where both SiO and \hcop{} are observed in absorption (column 3) and evaluated for individual Gaussian components (column 4). The value listed for 3C111A is superscritped with a $\dagger$ to indicate that this result was calculated using H$^{13}$CO$^+$ (the result, $0.18\times10^{-10}$ is the same if we use \hcop{}).}
    \label{tab:XSiO_no_Gauss}
\end{table*}

\subsection{Previous observations} \label{subsec:previous_obs}
Several of these sightlines have been observed in \sio{} and \hcop{} absorption previously with lower sensitivity.
\citet{SiO2000} searched for \sio{} absorption in the direction of both 3C111 ($\sigma_{\tau}\approx0.007$) and NRAO 530 ($\sigma_{\tau}\approx0.003$) at $\sim0.5~\kms{}$ velocity resolution. They reported significant detections in the integrated \sio{} optical depth for both sources, but their \sio{} absorption spectrum in the direction of 3C111 showed no obvious peaks and could not be easily decomposed into Gaussian components. Moreover, more recent work by \citet{Liszt2014} did not detect \sio{} absorption in the direction of 3C111 with better sensitivity ($\sigma_{\tau}\approx0.005$), albeit at lower velocity resolution ($\sim0.6~\kms{}$).
\cite{SiO2000} reported a significant detection of SiO absorption in the direction of NRAO 530 around $5~\kms{}$, where we find a marginal detection. SiO toward NRAO530 was later confirmed with ALMA by \citet{Ando2016}, though with only $3.4~\kms{}$ velocity resolution. \citet{Liszt2014} detected \sio{} toward BL Lac, but they did not discuss this result. 

Earlier, \citet{Turner1998} detected SiO in emission with very narrow linewidths ($<1\kms{}$) toward translucent molecular clouds.
Yet, emission observations with single-dish telescopes have lower angular resolution relative to absorption (interferometric) studies and are less accurate in constraining SiO column density. The beam size in \citet{Turner1998} was $\sim74^{\prime\prime}$, whereas we measure absorption against background continuum (point) sources unresolved at sub-arcsecond angular resolution.
In addition, the structures observed by \citet{Turner1998}
all have estimated densities $>10^{3}~\percc{}$ and visual extinctions $\gtrsim3~\rm{mag}$ \citep[where the visual extinction in that work, taken from model predictions for individual absorbing structures, represents a strict lower limit to the definition of visual extinction reported here in Table \ref{tab:sources}, which is taken directly from observations for the total line of sight;][]{Planck2016_GNILC}. As noted in \cite{SiO2000}, the translucent structures probed by \citet{Turner1998} in narrow-line \sio{} emission are probably significantly denser than most of the structures probed by the narrow-line \sio{} absorption in directions with $A_V\lesssim~\rm{few~mag}$ in this work. Moreover, four of the six directions where \citet{Turner1998} detected SiO emission were at $|b|\leq3.5^\circ$, whereas four of the five directions where we detect SiO in absorption are at $|b|>6^\circ$.

\hcop{} has been studied extensively in surveys of molecular absorption in the diffuse ISM. 3C111, NRAO 530, BL Lac, and 3C418 have previously been observed in \hcop{} absorption \citep{deGeus1996,LucasLiszt1996}. \hcop{} absorption profiles are stable over periods of a few to a few tens of years \citep{WilkindCombes1997,LisztLucas2000,Han2017,Ryb1_2022}. Our absorption profiles are consistent with previous observations in these directions, and are among the most sensitive in any of these directions.

\section{Localizing absorbing structures} \label{sec:environments}
We use two recently-constructed 3D dust maps \citep{Lallement2019,Leike2020} to attempt to identify the specific gas structures we detect in absorption (Section \ref{subsec:localizing_structures}). Given the spatial resolution ($1$--$5~\rm{pc}$), we do not consider the volume densities reported in these maps, but instead only use these maps to identify absorbing structures along the line of sight. 
Figure \ref{fig:Sources} shows the directions we observed relative to prominent Galactic structures (left panel), as well as cross-sections through the 3D dust maps in the direction of each background source. Table \ref{tab:densities} lists the distance at which the peak density is measured in each map (see discussion below). After using these dust maps---as well as \hi{} \citep{HI4PI} and CO \citep{DameCO,Dame22}---to identify the absorbing structures, we also find estimates from the literature for the true volume densities of specific absorbing clouds (Section \ref{subsec:densities}; Table \ref{tab:densities}). We discuss these observations on a source-by-source basis below.

\subsection{Distance constraints using 3D dust data} \label{subsec:localizing_structures}

\subsubsection{3C111} \label{subsubsec:3C111A}
We see two \hcop{} absorption components in the direction of 3C111, both associated with the California molecular cloud. California has an exceptionally low star formation rate for a cloud its size, and this sightline is not near any known YSOs \citep{Lada2010,Lada2017}, which can drive powerful shocks.

\begin{figure*}
    \centering
    \includegraphics[width=0.8\linewidth]{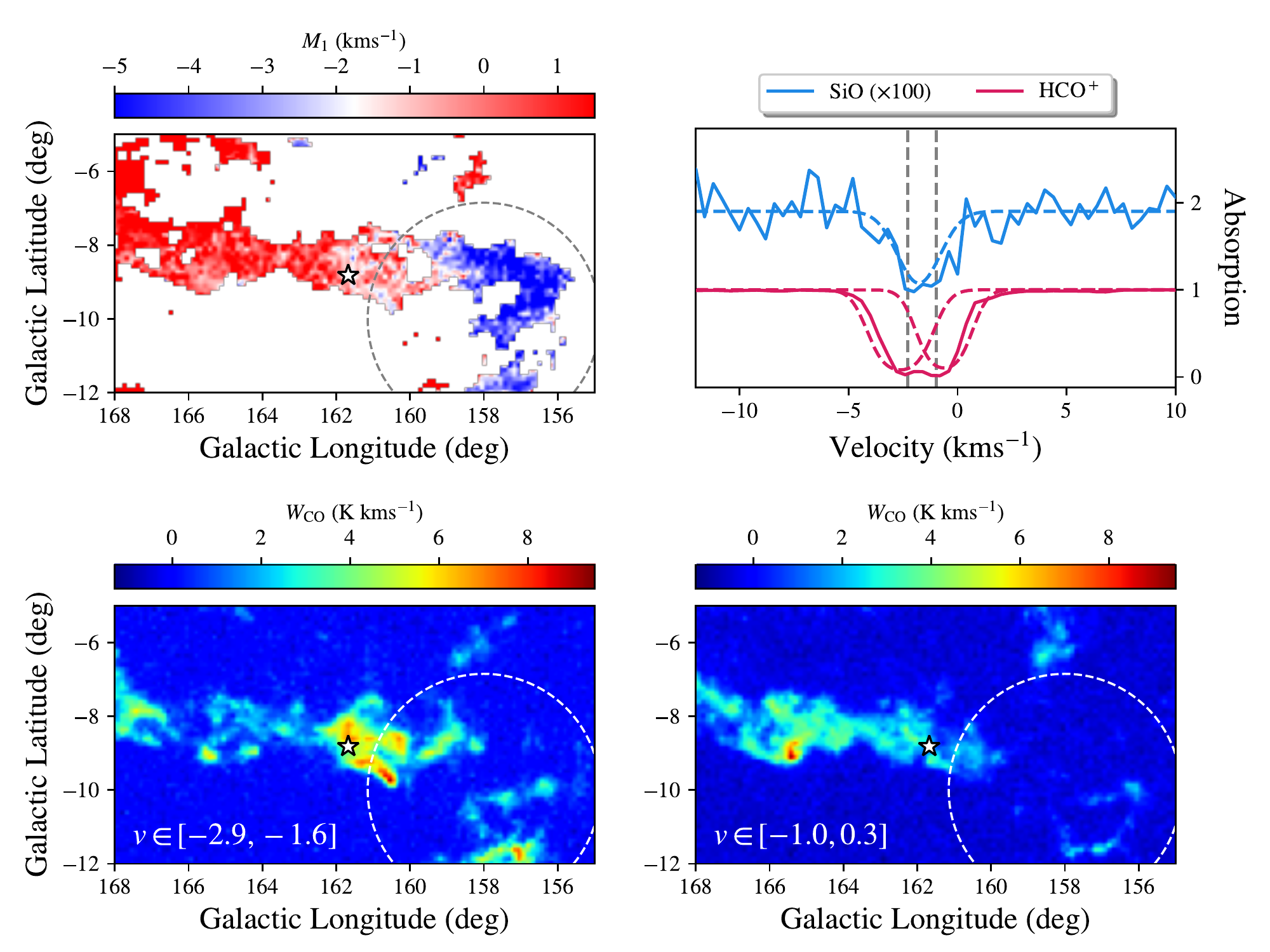}
    \caption{\textit{Top left:} Moment 1 map of the California molecular cloud seen in CO emission \citep{DameCO}. \textit{Top right:} \sio{} and \hcop{} absorption spectra ($e^{-\tau}$) in the direction of 3C111A. The SiO spectrum has been amplified by a factor of 100. \textit{Bottom left:} Integrated CO emission from $-2.9~\kms{}$ to $-1.6~\kms{}$ \citep{DameCO}, the interval roughly coincident with one of our \hcop{} absorption components and our \sio{} absorption component. \textit{Bottom right:} Integrated CO emission from $-1.0~\kms{}$ to $0.3~\kms{}$ \citep{DameCO}, the interval roughly coincident with the other \hcop{} absorption component. In the Moment 1 and integrated emission maps, the rough boundary of the bubble \citep{Rezaei2022} in California is shown as a dashed line. The position of 3C111A is shown as a white star. The $\sim-2.5~\kms{}$ component appears to be mainly associated with this eastern bubble component of the California molecular cloud, whereas the CO emission associated with the lower-velocity components is mainly associated with the western sheet-like component of the cloud.}
    \label{fig:3C111_env}
\end{figure*}

As confirmed in the detailed 3D map of California recently constructed by \citet{Rezaei2022}, this cloud complex comprises an extended sheet-like structure and a bubble $\sim25~\rm{pc}$ in radius, which lies at the eastern edge of the molecular cloud. We find that the $-2.5~\kms{}$ feature is associated with the shell of the eastern component, while the $-1~\kms{}$ feature is associated with the western sheet-like component. This is evident in the moment map and channel maps shown in Figure \ref{fig:3C111_env}, which show the boundary of the shell identified by \citet{Rezaei2022}. In the density profile from the \citet{Lallement2019} map, there are two clear peaks around California in this direction---the eastern (bubble) component is nearer, at a distance of $\sim450~\rm{pc}$, while the western (sheet-like) component is farther, at a distance of $\sim500~\rm{pc}$. The \sio{} absorption that we fit appears to be associated with the eastern bubble component around $-2.5~\kms{}$. We note that it is possible to derive a two-component fit to the \sio{} absorption with residuals similar to those for the single-component fit. In this case, we would have \sio{} associated with both the bubble and the sheet components of California. The estimate of $X_{\sio{}}$ would change only by a factor of $\sim2$ in this case from the single-component fit.

\subsubsection{2023+335}
Being our lowest-latitude background source, the sightline toward 2023+335 is very complex. We detect 7 components in \hcop{} absorption and 4 components in \sio{} absorption. Nearly all of this absorption is detected between $-5~\kms{}$ and $10~\kms{}$, associated with the Cygnus molecular cloud complex. As is evident in Figure \ref{fig:Sources}, there are many density peaks along the line of sight. Determining the distances to all of the structures seen in \hcop{} absorption is complex and beyond the scope of this work, but most of this absorption lies between 700--1500 pc, as indicated with a semi-transparent band in Figure \ref{fig:Sources}.

\subsubsection{3C418} \label{subsubsec:3C418}
There are 9 \hcop{} absorption components in the direction of 3C418 with velocities $-20~\kms{}\lesssim v \lesssim 20~\kms{}$. Most of these are likely associated with Cygnus---this sightline intersects the outskirts of the OB association Cyg OB7---although there are many density peaks along the line of sight, and determining the distances to all of the structures seen in \hcop{} absorption is complex and beyond the scope of this work. No \sio{} is detected at these velocities. 

Meanwhile, the $-57~\kms{}$ feature---where we detect a single component of both \hcop{} and \sio{}---is easily identified with the Perseus arm of the Galaxy at a distance of $\sim1.9~\rm{kpc}$. 
If we isolate the higher-velocity gas in this direction, we can estimate $A_V$ for the Perseus arm, $(N(\hi{})_{\rm{Per}}+2N(\htwo{})_{\rm{Per}})/(2.08\times10^{21}~\rm{cm^{-2}/mag})$ \citep{Zhu2017}, where $N(\hi{})_{\rm{Per}}$ and $N(\htwo{})_{\rm{Per}}$ are the \hi{} and \htwo{} (derived from \hcop{}) column densities calculated for the velocity interval including Perseus. Taking an extremely conservative interval, say, $-150~\kms{}\lesssim v\lesssim-45~\kms{}$,  we find $A_{V}\lesssim0.63$ for the Perseus arm.  This is a strict upper limit---if we take a tighter interval that seems to trace the gas around our \hcop{} and \sio{} absorption, say, $-75~\kms{}\lesssim v\lesssim-45~\kms{}$, we find $A_V\approx0.34$. This is the most diffuse environment where we detect \sio{}, and, to our knowledge, the most diffuse environment where \sio{} has been detected in the ISM.  

\citet{deGeus1996} searched for the $(1-0)$ lines of both \hcop{} and CO in absorption toward 3C418, detecting \hcop{} but not CO (the CO absorption spectrum from  had a sensitivity $\sigma_{\tau}\approx0.2$ at $0.32~\kms{}$ velocity resolution). CO is not detected in emission at these velocities in the survey of \citet{DameCO} ($\sigma_{T_B}\approx0.2~\rm{K}$).

\subsubsection{BL Lac}
\begin{figure*}
    \centering
    \includegraphics[width=0.93\linewidth]{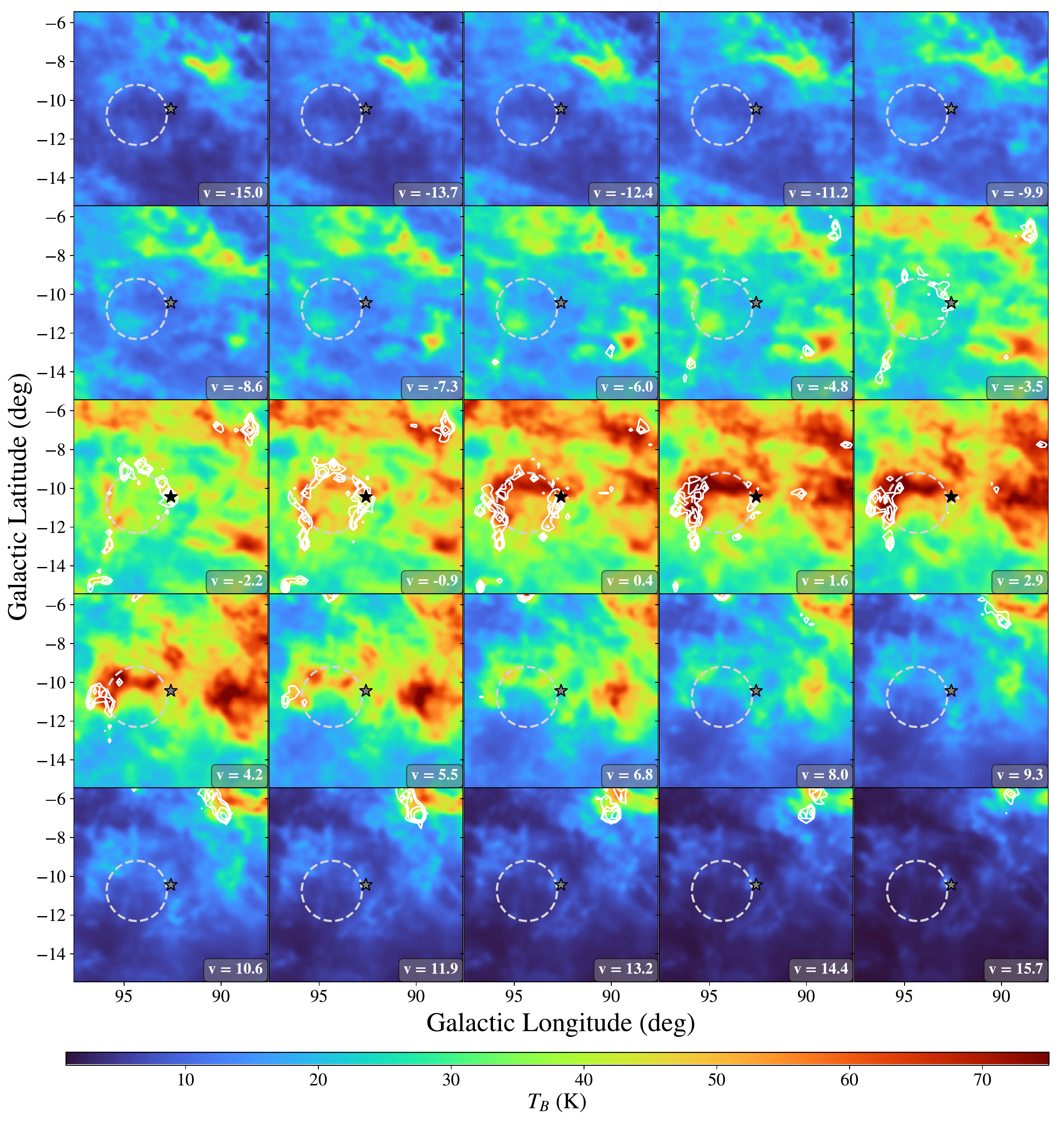}
    \caption{Maps of the \hi{} emission \citep{HI4PI} in the direction of BL Lac. White contours of the CO emission\cite{Dame22} are shown at $0.25~\rm{K}$ ,$0.5~\rm{K}$, $1~\rm{K}$, and $2~\rm{K}$. The observed velocity of the \hi{} and CO is indicated in the bottom right corner of each panel. The position of BL Lac is indicated with a star, colored black and gray for velocities with and without detected \hcop{} absorption, respectively. The approximate boundary of the shell
    discussed in the text is shown as a dashed gray circle in all panels.}
    \label{fig:BLLac_HICO}
\end{figure*}

BL Lac is in the direction of the Lacerta molecular cloud, but not particularly near Lac OB1 or LBN 437, which is modified by Lac OB1 \citep{ChenLee2008}.
The direction toward BL Lac intersects the eastern part of an arc-like structure associated with Lacerta. As shown in Figure \ref{fig:BLLac_HICO}, this arc is visible in both the dust \citep[e.g.,][]{Planck2016_GNILC} and CO \citep{Dame22}. It is clear from an inspection of the \citet{Lallement2019} 3D dust map that the emission north of $b\approx-12^\circ$ in this arc is associated with a single, coherent structure at a distance of $\sim530~\rm{pc}$---we interpret this as a shell $\sim10\rm{pc}$ in radius.
In Figure \ref{fig:BLLac_HICO}, we see that the \hi{} in the velocity interval where we detect \hcop{} and \sio{} ($-5~\kms{}$ to $5~\kms{}$; Table \ref{tab:XSiO_no_Gauss}) obeys a similar arc-like morphology \citep{HI4PI}. The CO in this direction is filamentary and relatively dense \citep{LisztPety2012}, with a moderately high excitation temperature $\sim7~\rm{K}$ \citep{Liszt1998} compared to similar sightlines. It is possible that turbulent flows driven by the expansion of the shell could contribute to these characteristics as well as the formation of SiO in this direction.

\subsubsection{NRAO 530}
The sightline toward NRAO 530 intersects the outskirts of the Aquila molecular cloud, around a distance of $\sim200~\rm{pc}$. The velocity of the \hcop{} and \sio{} features detected in this direction are consistent with their association with the Aquila molecular cloud \citep{DameCO}. CO has been detected in emission (peak CO brightness temperature $\approx0.5\text{--}0.6~\rm{K}$) and absorption ($N(\rm{CO})\approx1.1\times10^{15}\persc{}$) in this direction\cite{Liszt1998}.

\subsubsection{4C+28.07}
We detect \hcop{} at $-6~\kms{}$ and $-1~\kms{}$ in the direction of 4C+28.07 (see Figure \ref{fig:Sources}). The velocities of these features likely suggest an association with Perseus and Taurus, respectively. In Figure \ref{fig:Sources}, it is clear that this sightline intersects the Per-Tau shell \citep{Bialy2021}, confirming the association. 
We find faint CO in this direction at $-5~\kms{}\lesssim v \lesssim 2~\kms{}$ \citep{Dame22}. 

\pagebreak
\subsubsection{4C09.57}
We detect four \hcop{} components in this direction. All peaks are associated with peaks in \hi{} emission, with the three highest-velocity components being associated with more diffuse \hi{} than the lowest-velocity component. We find no CO emission in this direction \citep{Planck2014,Dame22}. In the 3D dust map from \citet{Leike2020}, it is clear that this sightline passes through at least one bubble-like feature at a distance of around $100~\rm{pc}$, the far side of which accounts for the density peak at $\sim150~\rm{pc}$ and seems to be associated with the local bubble wall.

\subsubsection{J0423-0120}
We detect one \hcop{} absorption component in this direction and find no CO emission in this direction \citep{Planck2014,Dame22}. In the dust maps, there is a density peak at $\sim150~\rm{pc}$, perhaps associated with the local bubble wall.

\subsection{Density constraints} \label{subsec:densities}
We first attempt to find density estimates from the literature in the direction of our background sources. \citet{Araya2014} observed H$_2$CO in absorption in the directions of 3C111 and BL Lac. They argued that the excitation temperature of H$_2$CO implies an upper limit to the density of $n_{\rm{H}}<10^5\percc{}$ in these directions, while the relatively weak absorption by H$_2$CO compared to CO and \hcop{} suggests that the density probably exceeds $n_{\rm{H}}\gtrsim10^2\percc{}$.
Meanwhile, NRAO 530 lies in the outskirts of the Aquila molecular cloud. Denser clouds and molecular cores in Aquila to have densities $\sim10^4~\percc{}$ \citep{Levshakov2013}. Presumably the density toward NRAO 530 is significantly lower, as it is in the much more diffuse outskirts of Aquila. We therefore suggest $10^2~\percc{}\lesssim n_{\rm{H}}\lesssim 10^3 \persc{}$ \citep{SnowMcCall2006,Levshakov2013}.
The molecular absorption associated with the Perseus arm in the direction of 3C418, where we detect SiO, is CO-dark at a sensitivity of $\sim0.1~\rm{K}$ \citep{DameCO}. \citet{Busch2019} showed that CO-dark gas in the Perseus arm has densities $\lesssim200~\percc{}$. Given the low $A_V$ \citep{SnowMcCall2006} and the absence of CO absorption and emission \citep[see Figure 5 of][]{Busch2019}, we estimate that the density of the feature detected in \sio{} and \hcop{} absorption toward 3C418 is $n\lesssim200~\percc{}$. 
For other structures not explicitly discussed in previous works, we can use rough limits established for diffuse, translucent, and dense clouds \citep{SnowMcCall2006}: $n_{\rm{H}}\sim 10^4~\percc{}$ for the structures in the direction of 2023+335 and $100~\percc{}\lesssim n_{\rm{H}}\lesssim 500~\percc{}$ for structures in the direction of 4C+28.07, 4C09.57, and J0423-0120. 

Alternatively, we can approximate the density of an interstellar structure by assuming some path length $L$, 
\begin{equation} \label{eq:n}
    n \approx \frac{N(\hi{})+2N(\htwo{})}{L},
\end{equation}
where $N(\hi{})$ is the \hi{} column density and $N(\htwo{})$ is the \htwo{} column density. We can estimate $N(\htwo{})\approx N(\hcop{})/3\times10^{-9}$ (see Section \ref{subsubsec:XSiO}), where the \hcop{} column density is solved by integrating over the Gaussian feature (Equation \ref{eq:N}). $N(\hi{})$ is more difficult to evaluate for individual structures because decomposing \hi{} emission spectra into Gaussian components and matching those components to molecular absorption features is a complex process \citep[e.g.,][]{21-SPONGE,Ryb1_2022}. Nevertheless, we can estimate $N(\hi{})$ using $N(\hi{})=1.823\times10^{18}\int T_B dv$ and integrating over $\pm1~$FWHM from the line center for each structure detected in absorption.
Solving Equation \ref{eq:n} in this way, we find densities $n\sim20$--$1500~\percc{}$ for $L=1~\rm{pc}$. In this case, only 2/31 structures have $n>10^3~\percc{}$, while 27/31 structures have $n<500~\percc{}$. Of course, these density estimates scale with $1/L$, so would increase if we assume a smaller path length.
For example, if we set $L$ to the characteristic size of so-called ``CO hot spots,'' clumps of of enhanced CO emission in the diffuse/translucent ISM, then $L\sim0.5~\rm{pc}$ and our density estimates increase by a factor of 2, which means that our estimates still largely fall in the range $n\sim10^2$--$10^{3}~\percc{}.$

The local densities can also be constrained by the weakness of \hcop{} emission \citep[e.g.,][]{LucasLiszt1996,LisztLucas2000}. \hcop{} is detected in emission toward 3C111 and BL Lac at levels of $\sim0.04~\rm{K}$ and $\sim0.1~\rm{K}$, respectively \citep{LucasLiszt1996}. While we are not aware of \hcop{} emission observations in the direction of our other background sources, sightlines with similar (or even slightly higher) $A_V$ and integrated \hcop{} optical depths tend to show emission at $\lesssim0.05~\rm{K}$, if detected at all \citep{LucasLiszt1996}. While we do not have direct constraints on the kinetic temperature in most directions, the atomic hydrogen kinematically associated with \hcop{} and \sio{} absorption toward 3C111 and BL Lac is $\sim40$--$80~\rm{K}$, and more broadly temperatures of \hi{} structures kinematically associated with diffuse/translucent \hcop{} absorption are $\lesssim80~\rm{K}$ \citep{Ryb2_2022}.
Given the \hcop{} column densities (Figure \ref{fig:SiO_abundance}), optical depths (Table \ref{tab:gaussian_fits}), low excitation temperatures \citep{vanderTak2007,Godard2010,Luo2020}, narrow linewidths (Figure \ref{fig:FWHM_HCOp_SiO}, Table \ref{tab:gaussian_fits}), and the estimated kinetic temperatures \citep{Ryb1_2022}, the weakness of the \hcop{} emission ($\lesssim0.05~\rm{K}$) implies densities $\lesssim\rm{few}\times10^2~\percc{}$ in most cases in our sample \citep{vanderTak2007,Liszt2020}.

Taking into consideration the various approaches to estimating densities above, we put approximate limits on the densities in Table \ref{tab:densities}. These estimates are largely consistent, with typical densities $\sim10^2$--$10^3~\percc{}$.

\begin{table*}
    \centering
    \begin{tabular}{|l|c|c|c|}
    \hline
    Source &   $n_{\rm{max}}$ dist. (L20) &  $n_{\rm{max}}$ dist. (L19)  & $n_{\rm{est}}$\\
     &  pc & pc & \percc{} \\
    \hline
    3C111A &     276 &     505 & $10^2$--$\rm{few}\times10^3$ \\
    BLLac &   164 &    530 & $10^2$--$10^3$ \\
    NRAO530 &    199 &    185 & $<10^3$ \\
    4C09.57 &     147 &    160 & $\lesssim 500$ \\
    J0423-0120 &    144 &    165 &  $\lesssim500$ \\
    2023+335 &    27 &   1110 & $\lesssim10^3$ \\
    3C418 &     361 &    355 & $\lesssim10^3$ \\
    3C418 - Perseus &       -- &   1890 & $\lesssim 200$ \\
    4C+28.07 &   263 &      255 & $\lesssim 500$ \\
\hline
\end{tabular}
    \caption{Summary of the estimated densities of gas structures along the lines of sight observed in this work. For each background source (column 1), we list the distance of the density peak in the 3D dust maps from Lallement et al. (2019) (``L19'')\cite{Lallement2019} and Leike et al. (2020) (``L20'')\cite{Leike2020} (columns 2 and 3, respectively). Finally, we list the estimated range of densities $n_{\rm{est}}$ for absorbing structures along the line of sight based on available literature and observations.}
    \label{tab:densities}
\end{table*}

\subsection{Estimating the ages of shells in the \hi{} and CO} \label{subsec:shell_ages}
As discussed above and seen in Figures \ref{fig:3C111_env} and \ref{fig:BLLac_HICO}, we find shells in the CO (and \hi{} in the case of BL Lac) where the SiO absorption is detected. Here we estimate the ages of these shells. 

The age of an expanding shell is $t_{\rm{exp}}\approx\frac{2}{5} \frac{r}{v_{\rm{exp}}}$, where $v_{\rm{exp}}$ is the expansion velocity and the factor of $2/5$ is appropriate for supernovae \citep{Chevalier1994}. The California shell, observed here in the direction of 3C111, has $r_{\rm{Cal}}=25~\rm{pc}$. The Lacerta shell, observed here in the direction of BL Lac, has $r_{\rm{Lac}}=10~\rm{pc}$. Because these sightlines are still within $\sim10^\circ$ of the Galactic plane, the \hi{} emission is bright at these velocities, and it is difficult to isolate any double-peaked emission signature of the shells in the \hi{} to identify an expansion velocity \citep{Ehlerova2005}. If we assume a late-evolution expansion velocity of $8~\rm{kms^{-1}}$ \citep[e.g.,][]{Ehlerova2005}, we find expansion times of $1.2~\rm{Myr}$ and $0.5~\rm{Myr}$, respectively. We can also consider a broader range of expansion velocities, as expansion velocities of \hi{} shells in the Outer Galaxy have been found to range from $\sim5~\kms{}$ to $\sim25~\kms{}$ \citep{Ehlerova2005}. In this case, we find that the shells seen toward 3C111 and BL Lac are $\sim0.4$--$2~\rm{Myr}$ and $\sim0.2$--$0.8~\rm{Myr}$ old, respectively.

\section{High SiO abundances in the diffuse ISM} \label{sec:results}
\begin{figure}
    \centering
    \includegraphics[width=3.5in]{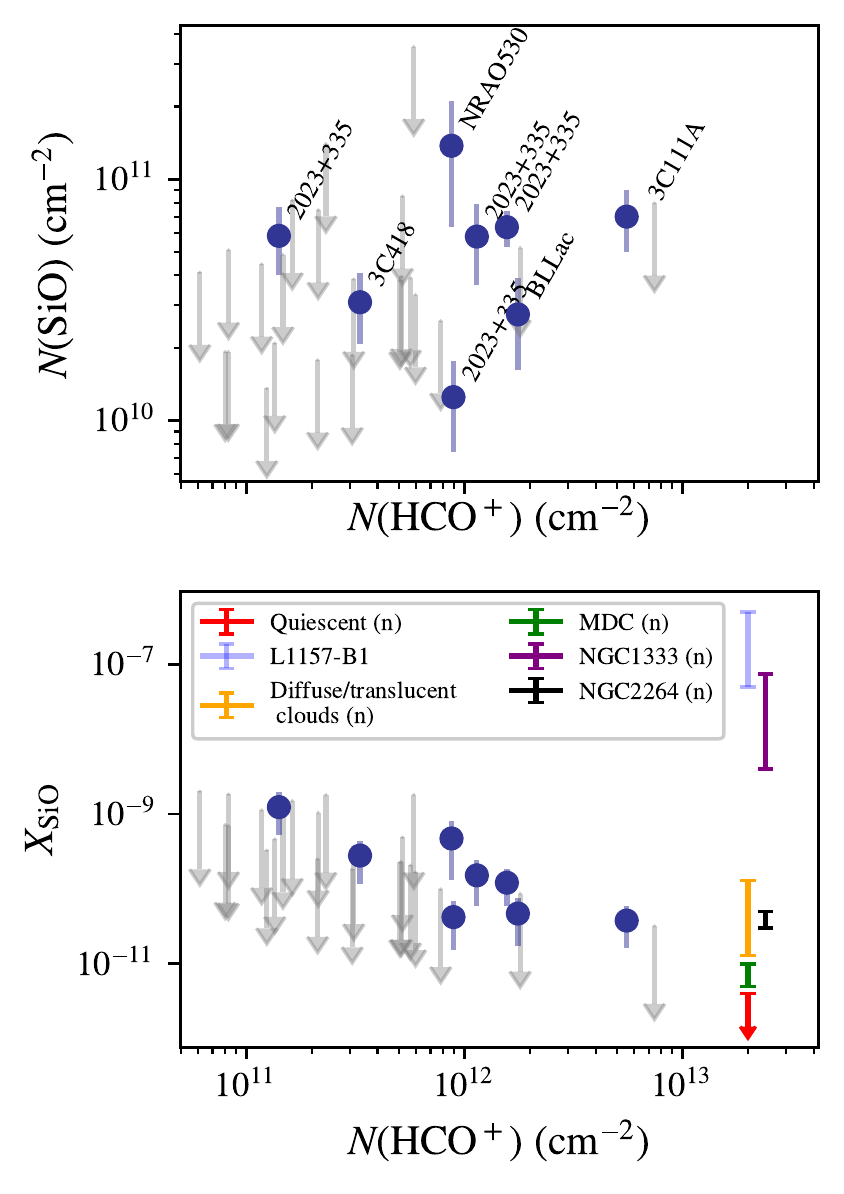}
    \caption{The SiO column density (top) and SiO abundance (bottom) versus the \hcop{} column density for components identified in our Gaussian decomposition. $3\sigma$ upper limits are shown in gray for non-detections. Detections shown in blue with $1\sigma$ error bars and are labeled in the top panel with the name of the background source against which each feature was detected (note the $x$-axes are the same in both panels). Previous SiO abundances measured in different interferometric experiments are shown for reference \citep{SiO2000,Spezzano2020,DuarteCabral2014,LopezSepulcre2016} (the $x$-axis values for these data are arbitrary). Results for narrow (FWHM$~\lesssim2~\kms{}$) SiO features are labeled with ``n'' in the legend.
    Uncertainties in $X_{\sio{}}$ include the uncertainty in $N(\hcop{})/N(\htwo{})$.
    We also show the estimated SiO abundance for quiescent gas measured in emission by \citet{MartinPintado1992} in red. 
    }
    \label{fig:SiO_abundance}
\end{figure}

In Figure \ref{fig:SiO_abundance}, we show the SiO column densities, $N(\sio{}) $ (top panel), and abundances, $X_{\sio{}}$
(bottom panel), for all components identified in the \sio{} absorption spectra. 
Where SiO is detected, $X_{\rm{SiO}}$ ranges from $3\times10^{-11}$ to $1\times10^{-9}$. 
Our key result is that in all cases
where we detect SiO, we find abundances a few to a few hundred times greater than those previously reported in quiescent environments, where $X_{\sio{}}\lesssim4\times10^{-12}$ \citep{MartinPintado1992}. 
Quiescent clouds, which show no star formation activity, lack shocks with high enough velocities ($\sim25$ km s$^{-1}$) to sputter Si off the grain cores. 
In Figure \ref{fig:SiO_abundance} we compare our results with the SiO abundances previously measured in high resolution interferometric observations \citep{SiO2000,DuarteCabral2014,Spezzano2020,DeSimone2022}, as well as the typically-assumed quiescent abundance \citep{MartinPintado1992}. 
Our abundances in diffuse and translucent directions are higher than what was found in non-star-forming environments, and almost as high as what was reported in dense star-forming regions with powerful outflows.

Such high SiO abundances are difficult to explain when considering
that our sightlines probe diffuse and translucent environments in the Milky Way: the three directions where we do not detect SiO have $A_V<0.5~\rm{mag}$; four of the directions where we detect SiO have $1~\rm{mag}<A_V<5~\rm{mag}$; only one direction where we detect SiO has $A_V>5~\rm{mag}$. 
Star formation in nearby clouds occurs only above a threshold of $A_V\sim7~\rm{mag}$ \citep{Lada2010}. All our directions probe regions with extinction well below $7~\rm{mag}$, suggesting that it is unlikely for recent or ongoing star formation to be responsible for the formation of SiO. 
Studies of star formation in translucent clouds are rare, but several of these high latitude clouds have no evidence of star formation \citep[e.g.,][]{Hearty1999,Brand2012}.
Moreover, the densities of the absorbing structures along these lines of sight are $n\sim10^2$--$10^3~\percc{}$ in a large majority of cases (Section \ref{sec:environments}).

In 2/5 directions where we detect SiO, the SiO appears to be associated with large, shell-like structures visible in CO (see Figures \ref{fig:3C111_env} and \ref{fig:BLLac_HICO}): the \sio{} in the direction of 3C111A appears to be associated with the bubble on the eastern side of the California molecular cloud \citep{Rezaei2022}, while the \sio{} in the direction of BL Lac is associated with a well-defined shell visible in both \hi{} and CO, associated with the Lacerta molecular cloud \citep[Figure \ref{fig:BLLac_HICO}][]{HI4PI,DameCO,Dame22}.
The shells around 3C111 and BL Lac do not appear particularly bright in H$\alpha$ emission \citep{WHAM2003,WHAM2010} or synchrotron or free-free emission \citep{Planck2016Results}---the gas inside these shells is not highly ionized. Moreover, the faint H$\alpha$ emission from diffuse and translucent structures could be largely caused by scattering off of dust particles rather than in situ ionized gas \citep{Mattila2007,Lehtinen2010}.

The absence of ongoing nearby star formation signatures in the directions of 3C111A and BL Lac suggests that the shell-like structures
are likely caused by a previous episode of star formation. 
Based on rough estimates of the expansion velocities, we suggest ages of $\sim0.2$--$2~\rm{Myr}$ for these shells. For comparison,
the \sio{} abundances in recently shocked gas around star forming regions are larger than $\sim10^{-8}$ \citep[e.g.,][]{JimenezSerra2010,Spezzano2020,DeSimone2022}, whereas the abundances measured toward BL Lac and 3C111 are $(3$--$5)\times10^{-11}$. Given the typical depletion time of SiO \citep{Bergin1998,Codella1999}, and especially considering the low densities observed here, it is plausible that the SiO absorption in these directions results from de-accelerated shocks in which the SiO has not yet been fully depleted/destroyed, or has been replenished through low velocity shocks \citep{NguyenLuong2013}.

Meanwhile, the remaining three sightlines with six \sio{} detections show no obvious shells/bubbles in the \hi{} and CO. 
NRAO 530 probes the far outskirts of the Aquila molecular cloud; 2023+335 intersects the Cygnus molecular cloud complex; and the \sio{} toward 3C418 is clearly associated with the Perseus arm. 
In these directions, we have no evidence for active (or recently, within few Myrs, active) star formation. 
For example, if we isolate the \hi{} and \htwo{} (by proxy of \hcop{}) from the Perseus arm in the direction of 3C418, we find an optical extinction $0.3 \lesssim A_V \lesssim 0.6$, making it the most diffuse region where \sio{} has ever been detected in the ISM \citep{SiO2000,Corby2018}. This is particularly interesting, as this feature has one of the highest \sio{} abundances in our sample, with $X_{\rm{SiO}}\approx2.8\times10^{-10}$. This is one of the highest \sio{} abundances ever detected in the diffuse/translucent ISM, and the highest detected where $A_V<1~\rm{mag}$. 
NRAO 530 and 2023+335, while at higher visual extinction, have similarly high \sio{} abundances. Even in the absence of star formation or any clearly evident expanding structures, the diffuse and translucent gas in these directions has remarkably high \sio{} abundances compared to the typically-assumed value in quiescent gas \citep{MartinPintado1992}. 

\begin{figure}
    \centering
    \includegraphics[width=3.5in]{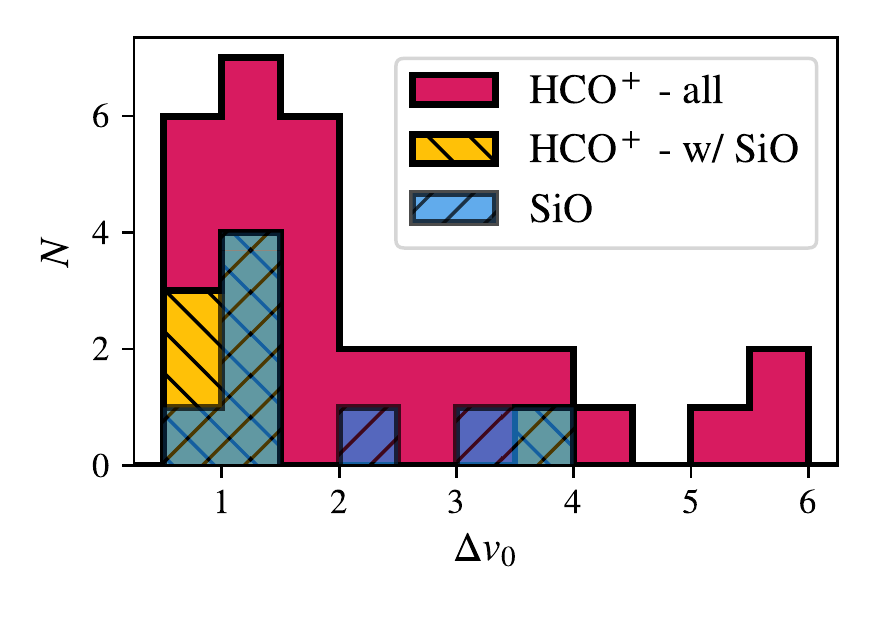}
    \caption{The distribution of full widths at half maximum (FWHMs) for Gaussian features identified in the \hcop{} absorption spectra and \sio{} absorption spectra. The unhatched dark pink histogram shows the distribution for all \hcop{} components; the diagonally hatched yellow histogram shows the distribution for \hcop{} components coincident with \sio{} absorption; the blue diagonally hatched histogram shows the distribution for all \sio{} components.}
    \label{fig:FWHM_HCOp_SiO}
\end{figure}

Figure \ref{fig:FWHM_HCOp_SiO} shows the distribution of full widths at half-maximum (FWHM) for both \hcop{} ($0.5~\kms{}<\rm{FWHM}<5.8~\kms{}$) and \sio{} ($0.8~\kms{}<\rm{FWHM}<3.9~\kms{}$) Gaussian absorption features.
Such narrow linewidths could allow us to exclude stellar outflows as a possible mechanism for SiO production. SiO lines are generally observed in protostellar outflows with a width of $20$--$50~\kms{}$ \citep{Gueth1998, Nisini07}. In principle, SiO from a post-shock layer could be associated with a linewidth of a few \kms{}. However this linewidth is found only in small regions of protostellar outflows (see, e.g., Fig. 7 of \citealt{Gueth1998} or Fig. 4 of \citealt{Leurini13}), and it is rather unlikely that we catch only this component with our experiment. In our case, the protostellar outflow origin is all the less likely that it is not compatible with the densities associated with our lines of sight (see Figure \ref{fig:Sources}, Table \ref{tab:densities}, and related discussions below).
We further note that the similarity in the line profiles of \hcop{} and \sio{} presented in Figure \ref{fig:spectra} and Table \ref{tab:gaussian_fits} could suggest that both species are formed under similar conditions \citep[e.g.,][]{Schilke2001}.

As shown in the top panel of Figure \ref{fig:SiO_abundance}, 
\sio{} column densities are relatively similar for structures with widely varying \hcop{} column densities. Under the assumption of a constant excitation temperature, the \sio{} column densities vary by a factor of 8, whereas the \hcop{} column densities vary by a factor of 38. 
This results in $X_{\sio{}}$ decreasing with increasing \hcop{} column density, as seen in the bottom panel of Figure \ref{fig:SiO_abundance}.
If we fit a power law $X(\rm{SiO})\propto N(\rm{HCO^+})^{\beta}$ to these points, we find $\beta=-1.29^{+0.48}_{-0.23}$, suggesting the decline is significant at a level $\sim 3\sigma$.

Six of the eight features with \sio{} detected have $N(\hcop{})\gtrsim10^{12}~\persc{}$, which is the column density at which \citet{Ryb2_2022} showed that
equilibrium chemical models fail to explain the observed column densities of several molecular species, including \hcop{}. In particular, \citet{Ryb2_2022} argued that non-equilibrium effects (i.e., shocks) are necessary to explain the \hcop{} column densities and \hi{} temperatures for features in the direction of 3C111A and BL Lac, where indeed we detect \sio{}, and where there appears to be an association with dynamical processes (shell expansion from a previous generation of star formation). Meanwhile, only one of the structures with $N(\hcop{})>10^{12}~\persc{}$ has no detected \sio{}\footnote{This is in the direction of 3C111A; see Section \ref{subsubsec:XSiO} and Table \ref{tab:XSiO_no_Gauss} for further discussion.}. 
These observations provide direct evidence that the high observed column densities of several molecular species (e.g., \hcop{}, HCN, C$_2$H, HNC) in the diffuse ISM, which cannot be explained with equilibrium chemistry, are associated with dynamical events and shocks that trigger non-equilibrium chemistry \citep{Godard2010,Ryb2_2022}.

\section{Discussion} \label{sec:discussion}
Our results suggest that the formation of SiO in the diffuse ISM (i.e., in the absence of significant star formation and stellar feedback) is more widespread and effective than previously reported \citep{MartinPintado1992}. Even quiescent environments with $n<10^{3}~\percc{}$ can produce SiO
with abundances similar to those found in much denser environments, with $n>10^4~\percc{}$ \citep[e.g.,][]{DuarteCabral2014,Louvet2016,DeSimone2022} or directions of protostellar outflows. 

While explaining the origin of low-velocity shocks and SiO abundances in dense environments has remained challenging, the SiO detections in diffuse regions that we report here are even harder to explain.
Densities probed by our observations are most likely in the range $10^2-10^3~\percc{}$ (Section \ref{sec:environments}). Our results therefore challenge the current understanding of how shocks process dust grains and the distribution and form of silicon in the quiescent medium. 
What physical processes can form low-velocity shocks in the diffuse ISM remains an open question, though in the cases of 3C111 and BL Lac, where shells are evident in the \hi{} and CO at velocities where we detect SiO, the SiO may be associated with expanding shells from old ($\sim 0.2\text{--}2~\rm{Myr}$) star formation episodes.

We suggest that the typically-assumed abundance of $X_{\sio{}}<4\times10^{-12}$ for quiescent environments from \citet{MartinPintado1992} likely needs to be reconsidered. 
For example, we have detected \sio{} in perhaps the most diffuse environment to date---from the Perseus arm in the direction of 3C418 where $n\sim200\percc{}$---which has an \sio{} abundance $X_{\sio{}}\approx2.8\times10^{-10}$, almost two orders of magnitude higher than the typically-assumed quiescent value.
Abundances derived from interferometric observations offer a better measure of the local conditions along the line of sight than those derived from single-dish observations because of the much smaller beam size. Our absorption-line observations against continuum sources trace a pencil-beam, offering the best available measure of $X_{\sio{}}$ in diffuse and translucent environments. Previous observations of \sio{} in environments with $n<10^3~\percc{}$ have been exceedingly rare; while \cite{SiO2000} hinted at elevated \sio{} abundances in diffuse and translucent environments, they did not investigate the abundances of individual structures (see Section \ref{subsec:gaussian_fits}) or find SiO at $A_V<1.5~\rm{mag}$. Our sample of \sio{} absorption from diffuse and translucent gas, the largest to date, clearly demonstrates that \sio{} abundances in these environments do not align with the \citet{MartinPintado1992} result.

It is possible that a larger quantity of Si is present free in the gas phase or in dust grain mantles in the diffuse ISM than previously thought and/or that SiO is formed largely in purely gas-phase reactions. 
Models show that the \sio{} column densities in low-velocity shocks are elevated by at least $\sim3$ orders of magnitude when $1$--$10\%$ of the pre-existing (pre-shock) Si is in the gas phase\footnote{In these models SiO is removed from the gas phase by conversion to SiO$_2$ or re-absorption onto the surface of grains on a somewhat long timescale, see discussion in \citet{Gusdorf2008a}.} compared to shocks where the Si is entirely locked in grain cores and mantles \citep[for $n\sim10^4\text{--}10^5~\percc$;][]{Gusdorf2008,NguyenLuong2013,Lesaffre2013,Louvet2016}. 
This suggested fraction of free Si from modeling is roughly consistent with observations (e.g., the heavy depletion case from \citealt{Jenkins2009}; see their Equations 1 and 2, and Table 4). Moreover, SiO production by weak shocks may depend on the history of the gas---previous strong shocks may have released Si into the gas phase or moved Si from the grain cores to the grain surfaces, therefore making it easier for a subsequent generation of weaker shocks to produce SiO\footnote{We note that the path $\mathrm{SiO_2}+\mathrm{H}\rightarrow \sio{}+\mathrm{OH}$ does not likely contribute to the formation of \sio{} in these environments because the reaction is highly endothermic and the conditions in which this reaction could take place are not likely to occur.}. In this scenario, in regions with no history of strong shocks, one could expect that the observed SiO abundance falls towards lower values, which could provide an interpretation for the upper limits in abundance shown in our Figure \ref{fig:SiO_abundance}.
Understanding the physical processes effective at releasing Si is essential for constraining dust grain destruction and accretion rates. We anticipate that mechanisms, such as cloud-cloud collisions, ram pressure, or turbulent dissipation---all of which can occur in the diffuse ISM---may play a crucial role in
processing of dust grains, and providing the initial conditions for 
the earliest stages of molecule formation.
These mechanisms need to be better understood and included in future numerical simulations of star formation and galaxy evolution.

Finally, the detection and high abundance of \sio{} in the diffuse and translucent environments observed here, especially in directions where there is no clear evidence for dynamical processes, may suggest that shocks are 
not the key vehicle for SiO production in at least some interstellar environments. 
For example, interstellar grains accelerated by MHD turbulence can be shattered \citep{Jones1996} in grain-grain collisions at high speeds through gyroresonance interactions even in regions that are not shocked \citep{Yan2004}. Grains reach sufficiently high velocities to shatter even in the relatively diffuse CNM \citep{Moseley2023}. The Si released when UV radiation destroys shattered grains could contribute to the formation of SiO in such environments, eliminating the need for shocks to uniquely explain the formation of SiO.
It has been argued that UV (and presumably X-rays and cosmic rays) can also release Si from grain cores \citep{Schilke2001}, and probably even more easily from grain mantles. The decrease in \sio{} with increasing \hcop{} column density may be in agreement with SiO originating in PDRs as a result of the UV radiation field \citep{Schilke2001}, since the thickness of the PDR layers does not depend on the total column density of the cloud. However, we expect relatively weak radiation fields in these directions compared to \citet{Schilke2001}, and \citet{Ryb2_2022} argued against PDRs (based on the \hi{} and \hcop{} properties) in at least two of the five directions where \sio{} is detected. With future measurements, we will provide a more robust test of whether $N(\sio{})$ remains similar across diverse environments \citep[Figure \ref{fig:SiO_abundance}; see also Figure 8 of][]{Schilke2001}.
While our results are important to clarify and inform the use of SiO as a shock tracer in the Milky Way and other galaxies, an understanding of additional chemical pathways effective in producing high SiO abundances purely from the gas-phase chemistry will require future studies and larger observational samples. Our future work will focus on a larger observational sample and detailed chemical modeling.

\section{Conclusions} \label{sec:conclusions}
We have observed \sio{} in absorption in the direction of 8 background continuum sources in diffuse and translucent directions. Though still modest, this represents the largest sample to date of SiO absorption in the diffuse and translucent ISM. These observations have important implications for the physics and chemsitry of diffuse gas in the Milky Way. 
\begin{enumerate}
    \item We detect \sio{} in 5/8 directions, including all directions with $A_V\gtrsim1$. For all structures detected in \sio{} absorption here, $A_V<5~\rm{mag}$, suggesting that these environments are all non-star-forming \citep{Lada2010}.
    \item The observed linewidths of \sio{} absorption lines are narrow ($0.8~\kms{}<\rm{FWHM}<3.9~\kms{}$), likely ruling out high-velocity shocks as a formation mechanism. 
    \item The \sio{} abundances we measure in the diffuse and translucent (non-star-forming) ISM are a few to a few hundred times greater than the typically assumed abundance in quiescent gas, $\lesssim4\times10^{-12}$ \citep{MartinPintado1992}.
    \item In at least two directions, \sio{} in the diffuse ISM is associated with shells visible in CO emission (and \hi{} emission in one case). We suggest that the \sio{} in these directions is associated with de-accelerated shocks from stellar feedback events $0.2$--$2~\rm{Myr}$ ago.
    \item \sio{} is detected in the two directions observed here where \citet{Ryb2_2022} showed that UV-dominated, equilibrium chemical models were unable to account for the high \hcop{} column densities. These are also the directions where shells are observed.
    \item We emphasize the need for observational constraints on the distribution of Si in the gas phase and grain mantles, which are crucial for understanding the physics of grain processing and chemistry in the ISM.
chemistry
\end{enumerate}

In this work, we have focused on observational constraints of the \sio{} abundance in the diffuse ISM. In upcoming work, we will use modeling to test the effectiveness of low-velocity shocks in the diffuse ISM in producing the \sio{} abundances observed here. Previous attempts to model low-velocity shocks have largely focused on denser environments \citep[e.g.,][]{NguyenLuong2013,Louvet2016} or have not directly modeled \sio{} emission or absorption \citep{Lesaffre2013}.
Future observations with NOEMA, ALMA and JWST will also be critical for understanding formation of SiO in both gas-phase and dust-enabled chemistry, as well as modeling the earliest stages of molecule formation, which take place in the diffuse ISM.

\vspace{0.5cm}
We are grateful to Benjamin Goddard, Bruce Draine, Guillaume Pineau des For\^ets, Ana L\'opez-Sepulcre, Bob Benjamin, J\'er\^ome Pety, and Maryvonne Gerin for helpful conversations about the interpretation of results presented here.
We thank the anonymous referee for constructive comments.
D.R. thanks Jan Martin Winters for support in the data reduction process. 
This work was supported by the University of Wisconsin--Madison Office of the Vice Chancellor for Research and Graduate Education with funding from the Wisconsin Alumni Research Foundation, the NSF Award AST-2108370 and the National Aeronautics and Space Administration under Grant No. 4200766703. Support for this work was provided by the NSF through award SOSPA6-030 from the NRAO.

\bibliography{refs}{}

\begin{thebibliography}{}
\expandafter\ifx\csname natexlab\endcsname\relax\def\natexlab#1{#1}\fi
\providecommand{\url}[1]{\href{#1}{#1}}
\providecommand{\dodoi}[1]{doi:~\href{http://doi.org/#1}{\nolinkurl{#1}}}
\providecommand{\doeprint}[1]{\href{http://ascl.net/#1}{\nolinkurl{http://ascl.net/#1}}}
\providecommand{\doarXiv}[1]{\href{https://arxiv.org/abs/#1}{\nolinkurl{https://arxiv.org/abs/#1}}}

\bibitem[{{Amo-Baladr{\'o}n} {et~al.}(2011){Amo-Baladr{\'o}n},
  {Mart{\'\i}n-Pintado}, \& {Mart{\'\i}n}}]{AmoBaladron2011}
{Amo-Baladr{\'o}n}, M.~A., {Mart{\'\i}n-Pintado}, J., \& {Mart{\'\i}n}, S.
  2011, \aap, 526, A54, \dodoi{10.1051/0004-6361/200913545}

\bibitem[{{Ando} {et~al.}(2016){Ando}, {Kohno}, {Tamura}, {Izumi}, {Umehata},
  \& {Nagai}}]{Ando2016}
{Ando}, R., {Kohno}, K., {Tamura}, Y., {et~al.} 2016, \pasj, 68, 6,
  \dodoi{10.1093/pasj/psv110}

\bibitem[{{Araya} {et~al.}(2014){Araya}, {Dieter-Conklin}, {Goss}, \&
  {Andreev}}]{Araya2014}
{Araya}, E.~D., {Dieter-Conklin}, N., {Goss}, W.~M., \& {Andreev}, N. 2014,
  \apj, 784, 129, \dodoi{10.1088/0004-637X/784/2/129}

\bibitem[{{Armijos-Abenda{\~n}o}
  {et~al.}(2020{\natexlab{a}}){Armijos-Abenda{\~n}o}, {Banda-Barrag{\'a}n},
  {Mart{\'\i}n-Pintado}, {D{\'e}nes}, {Federrath}, \&
  {Requena-Torres}}]{ArmijosAbendano2020}
{Armijos-Abenda{\~n}o}, J., {Banda-Barrag{\'a}n}, W.~E., {Mart{\'\i}n-Pintado},
  J., {et~al.} 2020{\natexlab{a}}, \mnras, 499, 4918,
  \dodoi{10.1093/mnras/staa3119}

\bibitem[{{Armijos-Abenda{\~n}o}
  {et~al.}(2020{\natexlab{b}}){Armijos-Abenda{\~n}o}, {Banda-Barrag{\'a}n},
  {Mart{\'\i}n-Pintado}, {D{\'e}nes}, {Federrath}, \&
  {Requena-Torres}}]{AA2020}
---. 2020{\natexlab{b}}, \mnras, 499, 4918, \dodoi{10.1093/mnras/staa3119}

\bibitem[{{Bergin} {et~al.}(1998){Bergin}, {Melnick}, \&
  {Neufeld}}]{Bergin1998}
{Bergin}, E.~A., {Melnick}, G.~J., \& {Neufeld}, D.~A. 1998, \apj, 499, 777,
  \dodoi{10.1086/305656}

\bibitem[{{Bialy} {et~al.}(2021){Bialy}, {Zucker}, {Goodman}, {Foley}, {Alves},
  {Semenov}, {Benjamin}, {Leike}, \& {En{\ss}lin}}]{Bialy2021}
{Bialy}, S., {Zucker}, C., {Goodman}, A., {et~al.} 2021, \apjl, 919, L5,
  \dodoi{10.3847/2041-8213/ac1f95}

\bibitem[{{Brand} {et~al.}(2012){Brand}, {Wouterloot}, \&
  {Magnani}}]{Brand2012}
{Brand}, J., {Wouterloot}, J.~G.~A., \& {Magnani}, L. 2012, \aap, 547, A85,
  \dodoi{10.1051/0004-6361/201219822}

\bibitem[{{Busch} {et~al.}(2019){Busch}, {Allen}, {Engelke}, {Hogg}, {Neufeld},
  \& {Wolfire}}]{Busch2019}
{Busch}, M.~P., {Allen}, R.~J., {Engelke}, P.~D., {et~al.} 2019, \apj, 883,
  158, \dodoi{10.3847/1538-4357/ab3a4b}

\bibitem[{{Chen} \& {Lee}(2008)}]{ChenLee2008}
{Chen}, W.~P., \& {Lee}, H.~T. 2008, in Handbook of Star Forming Regions,
  Volume I, ed. B.~{Reipurth}, Vol.~4, 124

\bibitem[{{Chevalier}(1994)}]{Chevalier1994}
{Chevalier}, R.~A. 1994, in Supernovae, ed. S.~A. {Bludman}, R.~{Mochkovitch},
  \& J.~{Zinn-Justin}, 743

\bibitem[{{Codella} {et~al.}(1999){Codella}, {Bachiller}, \&
  {Reipurth}}]{Codella1999}
{Codella}, C., {Bachiller}, R., \& {Reipurth}, B. 1999, \aap, 343, 585

\bibitem[{{Corby} {et~al.}(2018){Corby}, {McGuire}, {Herbst}, \&
  {Remijan}}]{Corby2018}
{Corby}, J.~F., {McGuire}, B.~A., {Herbst}, E., \& {Remijan}, A.~J. 2018, \aap,
  610, A10, \dodoi{10.1051/0004-6361/201730988}

\bibitem[{{Dame} {et~al.}(2001){Dame}, {Hartmann}, \& {Thaddeus}}]{DameCO}
{Dame}, T.~M., {Hartmann}, D., \& {Thaddeus}, P. 2001, \apj, 547, 792,
  \dodoi{10.1086/318388}

\bibitem[{Dame \& Thaddeus(2022)}]{Dame22}
Dame, T.~M., \& Thaddeus, P. 2022, The Astrophysical Journal Supplement Series,
  262, 5, \dodoi{10.3847/1538-4365/ac7e53}

\bibitem[{{de Geus} \& {Phillips}(1996)}]{deGeus1996}
{de Geus}, E.~J., \& {Phillips}, J.~A. 1996, in Unsolved Problems of the Milky
  Way, ed. L.~{Blitz} \& P.~J. {Teuben}, Vol. 169, 575

\bibitem[{{De Simone} {et~al.}(2022){De Simone}, {Codella}, {Ceccarelli},
  {L{\'o}pez-Sepulcre}, {Neri}, {Rivera-Ortiz}, {Busquet}, {Caselli},
  {Bianchi}, {Fontani}, {Lefloch}, {Oya}, \& {Pineda}}]{DeSimone2022}
{De Simone}, M., {Codella}, C., {Ceccarelli}, C., {et~al.} 2022, \mnras,
  \dodoi{10.1093/mnras/stac083}

\bibitem[{{Duarte-Cabral} {et~al.}(2014){Duarte-Cabral}, {Bontemps}, {Motte},
  {Gusdorf}, {Csengeri}, {Schneider}, \& {Louvet}}]{DuarteCabral2014}
{Duarte-Cabral}, A., {Bontemps}, S., {Motte}, F., {et~al.} 2014, \aap, 570, A1,
  \dodoi{10.1051/0004-6361/201423677}

\bibitem[{{Ehlerov{\'a}} \& {Palou{\v{s}}}(2005)}]{Ehlerova2005}
{Ehlerov{\'a}}, S., \& {Palou{\v{s}}}, J. 2005, \aap, 437, 101,
  \dodoi{10.1051/0004-6361:20034389}

\bibitem[{{Endres} {et~al.}(2016){Endres}, {Schlemmer}, {Schilke}, {Stutzki},
  \& {M{\"u}ller}}]{2016JMoSp.327...95E}
{Endres}, C.~P., {Schlemmer}, S., {Schilke}, P., {Stutzki}, J., \&
  {M{\"u}ller}, H. S.~P. 2016, Journal of Molecular Spectroscopy, 327, 95,
  \dodoi{10.1016/j.jms.2016.03.005}

\bibitem[{{Gerin} {et~al.}(2019){Gerin}, {Liszt}, {Neufeld}, {Godard},
  {Sonnentrucker}, {Pety}, \& {Roueff}}]{Gerin2019}
{Gerin}, M., {Liszt}, H., {Neufeld}, D., {et~al.} 2019, \aap, 622, A26,
  \dodoi{10.1051/0004-6361/201833661}

\bibitem[{{Gildas Team}(2013)}]{2013ascl.soft05010G}
{Gildas Team}. 2013, {GILDAS: Grenoble Image and Line Data Analysis Software}.
\newblock \doeprint{1305.010}

\bibitem[{{Godard} {et~al.}(2010){Godard}, {Falgarone}, {Gerin}, {Hily-Blant},
  \& {de Luca}}]{Godard2010}
{Godard}, B., {Falgarone}, E., {Gerin}, M., {Hily-Blant}, P., \& {de Luca}, M.
  2010, \aap, 520, A20, \dodoi{10.1051/0004-6361/201014283}

\bibitem[{{Godard} {et~al.}(2009){Godard}, {Falgarone}, \& {Pineau Des
  For{\^e}ts}}]{Godard2009}
{Godard}, B., {Falgarone}, E., \& {Pineau Des For{\^e}ts}, G. 2009, \aap, 495,
  847, \dodoi{10.1051/0004-6361:200810803}

\bibitem[{{Gong} {et~al.}(2020){Gong}, {Ostriker}, {Kim}, \& {Kim}}]{Gong2020}
{Gong}, M., {Ostriker}, E.~C., {Kim}, C.-G., \& {Kim}, J.-G. 2020, \apj, 903,
  142, \dodoi{10.3847/1538-4357/abbdab}

\bibitem[{{Gueth} {et~al.}(1998){Gueth}, {Guilloteau}, \&
  {Bachiller}}]{Gueth1998}
{Gueth}, F., {Guilloteau}, S., \& {Bachiller}, R. 1998, \aap, 333, 287

\bibitem[{{Gusdorf} {et~al.}(2008{\natexlab{a}}){Gusdorf}, {Cabrit}, {Flower},
  \& {Pineau Des For{\^e}ts}}]{Gusdorf2008}
{Gusdorf}, A., {Cabrit}, S., {Flower}, D.~R., \& {Pineau Des For{\^e}ts}, G.
  2008{\natexlab{a}}, \aap, 482, 809, \dodoi{10.1051/0004-6361:20078900}

\bibitem[{{Gusdorf} {et~al.}(2008{\natexlab{b}}){Gusdorf}, {Cabrit}, {Flower},
  \& {Pineau Des For{\^e}ts}}]{Gusdorf2008a}
---. 2008{\natexlab{b}}, \aap, 482, 809, \dodoi{10.1051/0004-6361:20078900}

\bibitem[{{Haffner} {et~al.}(2003){Haffner}, {Reynolds}, {Tufte}, {Madsen},
  {Jaehnig}, \& {Percival}}]{WHAM2003}
{Haffner}, L.~M., {Reynolds}, R.~J., {Tufte}, S.~L., {et~al.} 2003, \apjs, 149,
  405, \dodoi{10.1086/378850}

\bibitem[{{Haffner} {et~al.}(2010){Haffner}, {Reynolds}, {Madsen}, {Hill},
  {Barger}, {Jaehnig}, {Mierkiewicz}, {Percival}, \& {Chopra}}]{WHAM2010}
{Haffner}, L.~M., {Reynolds}, R.~J., {Madsen}, G.~J., {et~al.} 2010, in
  Astronomical Society of the Pacific Conference Series, Vol. 438, The Dynamic
  Interstellar Medium: A Celebration of the Canadian Galactic Plane Survey, ed.
  R.~{Kothes}, T.~L. {Landecker}, \& A.~G. {Willis}, 388.
\newblock \doarXiv{1008.0612}

\bibitem[{{Han} {et~al.}(2017){Han}, {Yun}, \& {Park}}]{Han2017}
{Han}, J., {Yun}, Y., \& {Park}, Y.-S. 2017, Journal of Korean Astronomical
  Society, 50, 185, \dodoi{10.5303/JKAS.2017.50.6.185}

\bibitem[{{Hatchell} {et~al.}(2001){Hatchell}, {Fuller}, \&
  {Millar}}]{Hatchell2001}
{Hatchell}, J., {Fuller}, G.~A., \& {Millar}, T.~J. 2001, \aap, 372, 281,
  \dodoi{10.1051/0004-6361:20010468}

\bibitem[{{Hearty} {et~al.}(1999){Hearty}, {Magnani}, {Caillault},
  {Neuh{\"a}user}, {Schmitt}, \& {Stauffer}}]{Hearty1999}
{Hearty}, T., {Magnani}, L., {Caillault}, J.~P., {et~al.} 1999, \aap, 341, 163

\bibitem[{{Herbst} {et~al.}(1989){Herbst}, {Millar}, {Wlodek}, \&
  {Bohme}}]{Herbst1989}
{Herbst}, E., {Millar}, T.~J., {Wlodek}, S., \& {Bohme}, D.~K. 1989, \aap, 222,
  205

\bibitem[{{HI4PI Collaboration} {et~al.}(2016){HI4PI Collaboration}, {Ben
  Bekhti}, {Fl{\"o}er}, {Keller}, {Kerp}, {Lenz}, {Winkel}, {Bailin},
  {Calabretta}, {Dedes}, {Ford}, {Gibson}, {Haud}, {Janowiecki}, {Kalberla},
  {Lockman}, {McClure-Griffiths}, {Murphy}, {Nakanishi}, {Pisano}, \&
  {Staveley-Smith}}]{HI4PI}
{HI4PI Collaboration}, {Ben Bekhti}, N., {Fl{\"o}er}, L., {et~al.} 2016, \aap,
  594, A116, \dodoi{10.1051/0004-6361/201629178}

\bibitem[{{Hopkins} {et~al.}(2012){Hopkins}, {Quataert}, \&
  {Murray}}]{Hopkins2012}
{Hopkins}, P.~F., {Quataert}, E., \& {Murray}, N. 2012, \mnras, 421, 3488,
  \dodoi{10.1111/j.1365-2966.2012.20578.x}

\bibitem[{{Jenkins}(2009)}]{Jenkins2009}
{Jenkins}, E.~B. 2009, \apj, 700, 1299, \dodoi{10.1088/0004-637X/700/2/1299}

\bibitem[{{Jim{\'e}nez-Serra} {et~al.}(2010){Jim{\'e}nez-Serra}, {Caselli},
  {Tan}, {Hernandez}, {Fontani}, {Butler}, \& {van Loo}}]{JimenezSerra2010}
{Jim{\'e}nez-Serra}, I., {Caselli}, P., {Tan}, J.~C., {et~al.} 2010, \mnras,
  406, 187, \dodoi{10.1111/j.1365-2966.2010.16698.x}

\bibitem[{{Jones} {et~al.}(1996){Jones}, {Tielens}, \&
  {Hollenbach}}]{Jones1996}
{Jones}, A.~P., {Tielens}, A.~G.~G.~M., \& {Hollenbach}, D.~J. 1996, \apj, 469,
  740, \dodoi{10.1086/177823}

\bibitem[{{Lada} {et~al.}(2017){Lada}, {Lewis}, {Lombardi}, \&
  {Alves}}]{Lada2017}
{Lada}, C.~J., {Lewis}, J.~A., {Lombardi}, M., \& {Alves}, J. 2017, \aap, 606,
  A100, \dodoi{10.1051/0004-6361/201731221}

\bibitem[{{Lada} {et~al.}(2010){Lada}, {Lombardi}, \& {Alves}}]{Lada2010}
{Lada}, C.~J., {Lombardi}, M., \& {Alves}, J.~F. 2010, \apj, 724, 687,
  \dodoi{10.1088/0004-637X/724/1/687}

\bibitem[{{Lallement} {et~al.}(2019){Lallement}, {Babusiaux}, {Vergely},
  {Katz}, {Arenou}, {Valette}, {Hottier}, \& {Capitanio}}]{Lallement2019}
{Lallement}, R., {Babusiaux}, C., {Vergely}, J.~L., {et~al.} 2019, \aap, 625,
  A135, \dodoi{10.1051/0004-6361/201834695}

\bibitem[{{Lehtinen} {et~al.}(2010){Lehtinen}, {Juvela}, \&
  {Mattila}}]{Lehtinen2010}
{Lehtinen}, K., {Juvela}, M., \& {Mattila}, K. 2010, \aap, 517, A79,
  \dodoi{10.1051/0004-6361/200913982}

\bibitem[{{Leike} {et~al.}(2020){Leike}, {Glatzle}, \&
  {En{\ss}lin}}]{Leike2020}
{Leike}, R.~H., {Glatzle}, M., \& {En{\ss}lin}, T.~A. 2020, \aap, 639, A138,
  \dodoi{10.1051/0004-6361/202038169}

\bibitem[{{Lesaffre} {et~al.}(2013){Lesaffre}, {Pineau des For{\^e}ts},
  {Godard}, {Guillard}, {Boulanger}, \& {Falgarone}}]{Lesaffre2013}
{Lesaffre}, P., {Pineau des For{\^e}ts}, G., {Godard}, B., {et~al.} 2013, \aap,
  550, A106, \dodoi{10.1051/0004-6361/201219928}

\bibitem[{{Leurini} {et~al.}(2013){Leurini}, {Codella}, {Gusdorf}, {Zapata},
  {G{\'o}mez-Ruiz}, {Testi}, \& {Pillai}}]{Leurini13}
{Leurini}, S., {Codella}, C., {Gusdorf}, A., {et~al.} 2013, \aap, 554, A35,
  \dodoi{10.1051/0004-6361/201118154}

\bibitem[{{Levshakov} {et~al.}(2013){Levshakov}, {Henkel}, {Reimers}, {Wang},
  {Mao}, {Wang}, \& {Xu}}]{Levshakov2013}
{Levshakov}, S.~A., {Henkel}, C., {Reimers}, D., {et~al.} 2013, \aap, 553, A58,
  \dodoi{10.1051/0004-6361/201220354}

\bibitem[{{Li} {et~al.}(2019){Li}, {Wang}, {Fang}, {Zhang}, {Li}, {Zhang},
  {Li}, {Zhu}, \& {Zeng}}]{Li2019}
{Li}, S., {Wang}, J., {Fang}, M., {et~al.} 2019, \apj, 878, 29,
  \dodoi{10.3847/1538-4357/ab1e4c}

\bibitem[{{Liszt} \& {Gerin}(2023)}]{LisztGerin2023}
{Liszt}, H., \& {Gerin}, M. 2023, \apj, 943, 172,
  \dodoi{10.3847/1538-4357/acae83}

\bibitem[{{Liszt} {et~al.}(2018){Liszt}, {Gerin}, \& {Grenier}}]{Liszt2018}
{Liszt}, H., {Gerin}, M., \& {Grenier}, I. 2018, \aap, 617, A54,
  \dodoi{10.1051/0004-6361/201833167}

\bibitem[{{Liszt} \& {Lucas}(2000)}]{LisztLucas2000}
{Liszt}, H., \& {Lucas}, R. 2000, \aap, 355, 333

\bibitem[{{Liszt} \& {Lucas}(2001)}]{Liszt2001HCN}
---. 2001, \aap, 370, 576, \dodoi{10.1051/0004-6361:20010260}

\bibitem[{{Liszt}(2020)}]{Liszt2020}
{Liszt}, H.~S. 2020, \apj, 897, 104, \dodoi{10.3847/1538-4357/ab9601}

\bibitem[{{Liszt} \& {Lucas}(1998)}]{Liszt1998}
{Liszt}, H.~S., \& {Lucas}, R. 1998, \aap, 339, 561

\bibitem[{{Liszt} \& {Pety}(2012)}]{LisztPety2012}
{Liszt}, H.~S., \& {Pety}, J. 2012, \aap, 541, A58,
  \dodoi{10.1051/0004-6361/201218771}

\bibitem[{{Liszt} {et~al.}(2014){Liszt}, {Pety}, {Gerin}, \&
  {Lucas}}]{Liszt2014}
{Liszt}, H.~S., {Pety}, J., {Gerin}, M., \& {Lucas}, R. 2014, \aap, 564, A64,
  \dodoi{10.1051/0004-6361/201323320}

\bibitem[{{Liszt} {et~al.}(2010){Liszt}, {Pety}, \& {Lucas}}]{Liszt2010}
{Liszt}, H.~S., {Pety}, J., \& {Lucas}, R. 2010, \aap, 518, A45,
  \dodoi{10.1051/0004-6361/201014510}

\bibitem[{{L{\'o}pez-Sepulcre} {et~al.}(2016){L{\'o}pez-Sepulcre}, {Watanabe},
  {Sakai}, {Furuya}, {Saruwatari}, \& {Yamamoto}}]{LopezSepulcre2016}
{L{\'o}pez-Sepulcre}, A., {Watanabe}, Y., {Sakai}, N., {et~al.} 2016, \apj,
  822, 85, \dodoi{10.3847/0004-637X/822/2/85}

\bibitem[{{Louvet} {et~al.}(2016){Louvet}, {Motte}, {Gusdorf}, {Nguy{\^e}n
  Luong}, {Lesaffre}, {Duarte-Cabral}, {Maury}, {Schneider}, {Hill}, {Schilke},
  \& {Gueth}}]{Louvet2016}
{Louvet}, F., {Motte}, F., {Gusdorf}, A., {et~al.} 2016, \aap, 595, A122,
  \dodoi{10.1051/0004-6361/201629077}

\bibitem[{{Lucas} \& {Liszt}(1996)}]{LucasLiszt1996}
{Lucas}, R., \& {Liszt}, H. 1996, \aap, 307, 237

\bibitem[{{Lucas} \& {Liszt}(1998)}]{LucasLiszt1998}
---. 1998, \aap, 337, 246

\bibitem[{{Lucas} \& {Liszt}(2000)}]{SiO2000}
{Lucas}, R., \& {Liszt}, H.~S. 2000, \aap, 355, 327

\bibitem[{{Luo} {et~al.}(2020){Luo}, {Li}, {Tang}, {Dawson}, {Dickey},
  {Bronfman}, {Qin}, {Gibson}, {Plambeck}, {Finger}, {Green}, {Mardones},
  {Koo}, \& {Lo}}]{Luo2020}
{Luo}, G., {Li}, D., {Tang}, N., {et~al.} 2020, \apjl, 889, L4,
  \dodoi{10.3847/2041-8213/ab6337}

\bibitem[{{Martin-Pintado} {et~al.}(1992){Martin-Pintado}, {Bachiller}, \&
  {Fuente}}]{MartinPintado1992}
{Martin-Pintado}, J., {Bachiller}, R., \& {Fuente}, A. 1992, \aap, 254, 315

\bibitem[{{Mart{\'\i}n-Pintado} {et~al.}(1997){Mart{\'\i}n-Pintado}, {de
  Vicente}, {Fuente}, \& {Planesas}}]{MartinPintado1997}
{Mart{\'\i}n-Pintado}, J., {de Vicente}, P., {Fuente}, A., \& {Planesas}, P.
  1997, \apjl, 482, L45, \dodoi{10.1086/310691}

\bibitem[{{Mattila} {et~al.}(2007){Mattila}, {Juvela}, \&
  {Lehtinen}}]{Mattila2007}
{Mattila}, K., {Juvela}, M., \& {Lehtinen}, K. 2007, \apjl, 654, L131,
  \dodoi{10.1086/511009}

\bibitem[{{McKee} \& {Ostriker}(1977)}]{McKeeOstriker1977}
{McKee}, C.~F., \& {Ostriker}, J.~P. 1977, \apj, 218, 148,
  \dodoi{10.1086/155667}

\bibitem[{{Moseley} {et~al.}(2023){Moseley}, {Teyssier}, \&
  {Draine}}]{Moseley2023}
{Moseley}, E.~R., {Teyssier}, R., \& {Draine}, B.~T. 2023, \mnras, 518, 2825,
  \dodoi{10.1093/mnras/stac3231}

\bibitem[{{M{\"u}ller} {et~al.}(2001){M{\"u}ller}, {Thorwirth}, {Roth}, \&
  {Winnewisser}}]{2001A&A...370L..49M}
{M{\"u}ller}, H.~S.~P., {Thorwirth}, S., {Roth}, D.~A., \& {Winnewisser}, G.
  2001, \aap, 370, L49, \dodoi{10.1051/0004-6361:20010367}

\bibitem[{{Murray} {et~al.}(2018){Murray}, {Stanimirovi{\'c}}, {Goss},
  {Heiles}, {Dickey}, {Babler}, \& {Kim}}]{21-SPONGE}
{Murray}, C.~E., {Stanimirovi{\'c}}, S., {Goss}, W.~M., {et~al.} 2018, \apjs,
  238, 14, \dodoi{10.3847/1538-4365/aad81a}

\bibitem[{{Nguyen-Lu'o'ng} {et~al.}(2013){Nguyen-Lu'o'ng}, {Motte}, {Carlhoff},
  {Louvet}, {Lesaffre}, {Schilke}, {Hill}, {Hennemann}, {Gusdorf}, {Didelon},
  {Schneider}, {Bontemps}, {Duarte-Cabral}, {Menten}, {Martin}, {Wyrowski},
  {Bendo}, {Roussel}, {Bernard}, {Bronfman}, {Henning}, {Kramer}, \&
  {Heitsch}}]{NguyenLuong2013}
{Nguyen-Lu'o'ng}, Q., {Motte}, F., {Carlhoff}, P., {et~al.} 2013, \apj, 775,
  88, \dodoi{10.1088/0004-637X/775/2/88}

\bibitem[{{Nisini} {et~al.}(2007){Nisini}, {Codella}, {Giannini}, {Santiago
  Garcia}, {Richer}, {Bachiller}, \& {Tafalla}}]{Nisini07}
{Nisini}, B., {Codella}, C., {Giannini}, T., {et~al.} 2007, \aap, 462, 163,
  \dodoi{10.1051/0004-6361:20065621}

\bibitem[{{Pety}(2005)}]{2005sf2a.conf..721P}
{Pety}, J. 2005, in SF2A-2005: Semaine de l'Astrophysique Francaise, ed.
  F.~{Casoli}, T.~{Contini}, J.~M. {Hameury}, \& L.~{Pagani}, 721

\bibitem[{{Planck Collaboration} {et~al.}(2014){Planck Collaboration}, {Ade},
  {Aghanim}, {Alves}, {Armitage-Caplan}, {Arnaud}, {Ashdown},
  {Atrio-Barandela}, {Aumont}, {Baccigalupi}, {Banday}, {Barreiro}, {Bartlett},
  {Battaner}, {Benabed}, {Beno{\^\i}t}, {Benoit-L{\'e}vy}, {Bernard},
  {Bersanelli}, {Bielewicz}, {Bobin}, {Bock}, {Bonaldi}, {Bond}, {Borrill},
  {Bouchet}, {Boulanger}, {Bridges}, {Bucher}, {Burigana}, {Butler}, {Cardoso},
  {Catalano}, {Chamballu}, {Chary}, {Chen}, {Chiang}, {Chiang}, {Christensen},
  {Church}, {Clements}, {Colombi}, {Colombo}, {Combet}, {Couchot}, {Coulais},
  {Crill}, {Curto}, {Cuttaia}, {Danese}, {Davies}, {de Bernardis}, {de Rosa},
  {de Zotti}, {Delabrouille}, {Delouis}, {Dempsey}, {D{\'e}sert}, {Dickinson},
  {Diego}, {Dole}, {Donzelli}, {Dor{\'e}}, {Douspis}, {Dupac}, {Efstathiou},
  {En{\ss}lin}, {Eriksen}, {Falgarone}, {Finelli}, {Forni}, {Frailis},
  {Franceschi}, {Fukui}, {Galeotta}, {Ganga}, {Giard}, {Giraud-H{\'e}raud},
  {Gonz{\'a}lez-Nuevo}, {G{\'o}rski}, {Gratton}, {Gregorio}, {Gruppuso},
  {Handa}, {Hansen}, {Hanson}, {Harrison}, {Henrot-Versill{\'e}},
  {Hern{\'a}ndez-Monteagudo}, {Herranz}, {Hildebrandt}, {Hily-Blant}, {Hivon},
  {Hobson}, {Holmes}, {Hornstrup}, {Hovest}, {Huffenberger}, {Hurier}, {Jaffe},
  {Jaffe}, {Jewell}, {Jones}, {Juvela}, {Keih{\"a}nen}, {Keskitalo}, {Kisner},
  {Knoche}, {Knox}, {Kunz}, {Kurki-Suonio}, {Lagache}, {L{\"a}hteenm{\"a}ki},
  {Lamarre}, {Lasenby}, {Laureijs}, {Lawrence}, {Leonardi}, {Le{\'o}n-Tavares},
  {Lesgourgues}, {Liguori}, {Lilje}, {Linden-V{\o}rnle}, {L{\'o}pez-Caniego},
  {Lubin}, {Mac{\'\i}as-P{\'e}rez}, {Maffei}, {Mandolesi}, {Maris}, {Marshall},
  {Martin}, {Mart{\'\i}nez-Gonz{\'a}lez}, {Masi}, {Massardi}, {Matarrese},
  {Matthai}, {Mazzotta}, {McGehee}, {Melchiorri}, {Mendes}, {Mennella},
  {Migliaccio}, {Mitra}, {Miville-Desch{\^e}nes}, {Moneti}, {Montier}, {Moore},
  {Morgante}, {Morino}, {Mortlock}, {Munshi}, {Murphy}, {Nakajima}, {Naselsky},
  {Nati}, {Natoli}, {Netterfield}, {N{\o}rgaard-Nielsen}, {Noviello},
  {Novikov}, {Novikov}, {Okuda}, {Osborne}, {Oxborrow}, {Paci}, {Pagano},
  {Pajot}, {Paladini}, {Paoletti}, {Pasian}, {Patanchon}, {Perdereau},
  {Perotto}, {Perrotta}, {Piacentini}, {Piat}, {Pierpaoli}, {Pietrobon},
  {Plaszczynski}, {Pointecouteau}, {Polenta}, {Ponthieu}, {Popa}, {Poutanen},
  {Pratt}, {Pr{\'e}zeau}, {Prunet}, {Puget}, {Rachen}, {Reach}, {Rebolo},
  {Reinecke}, {Remazeilles}, {Renault}, {Ricciardi}, {Riller}, {Ristorcelli},
  {Rocha}, {Rosset}, {Roudier}, {Rowan-Robinson}, {Rubi{\~n}o-Mart{\'\i}n},
  {Rusholme}, {Sandri}, {Santos}, {Savini}, {Scott}, {Seiffert}, {Shellard},
  {Spencer}, {Starck}, {Stolyarov}, {Stompor}, {Sudiwala}, {Sunyaev}, {Sureau},
  {Sutton}, {Suur-Uski}, {Sygnet}, {Tauber}, {Tavagnacco}, {Terenzi}, {Thomas},
  {Toffolatti}, {Tomasi}, {Torii}, {Tristram}, {Tucci}, {Tuovinen}, {Umana},
  {Valenziano}, {Valiviita}, {Van Tent}, {Vielva}, {Villa}, {Vittorio}, {Wade},
  {Wandelt}, {Wehus}, {Yamamoto}, {Yoda}, {Yvon}, {Zacchei}, \&
  {Zonca}}]{Planck2014}
{Planck Collaboration}, {Ade}, P.~A.~R., {Aghanim}, N., {et~al.} 2014, \aap,
  571, A13, \dodoi{10.1051/0004-6361/201321553}

\bibitem[{{Planck Collaboration} {et~al.}(2016{\natexlab{a}}){Planck
  Collaboration}, {Aghanim}, {Ashdown}, {Aumont}, {Baccigalupi}, {Ballardini},
  {Banday}, {Barreiro}, {Bartolo}, {Basak}, {Benabed}, {Bernard}, {Bersanelli},
  {Bielewicz}, {Bonavera}, {Bond}, {Borrill}, {Bouchet}, {Boulanger},
  {Burigana}, {Calabrese}, {Cardoso}, {Carron}, {Chiang}, {Colombo}, {Comis},
  {Couchot}, {Coulais}, {Crill}, {Curto}, {Cuttaia}, {de Bernardis}, {de
  Zotti}, {Delabrouille}, {Di Valentino}, {Dickinson}, {Diego}, {Dor{\'e}},
  {Douspis}, {Ducout}, {Dupac}, {Dusini}, {Elsner}, {En{\ss}lin}, {Eriksen},
  {Falgarone}, {Fantaye}, {Finelli}, {Forastieri}, {Frailis}, {Fraisse},
  {Franceschi}, {Frolov}, {Galeotta}, {Galli}, {Ganga}, {G{\'e}nova-Santos},
  {Gerbino}, {Ghosh}, {Giraud-H{\'e}raud}, {Gonz{\'a}lez-Nuevo}, {G{\'o}rski},
  {Gruppuso}, {Gudmundsson}, {Hansen}, {Helou}, {Henrot-Versill{\'e}},
  {Herranz}, {Hivon}, {Huang}, {Jaffe}, {Jones}, {Keih{\"a}nen}, {Keskitalo},
  {Kiiveri}, {Kisner}, {Krachmalnicoff}, {Kunz}, {Kurki-Suonio}, {Lamarre},
  {Langer}, {Lasenby}, {Lattanzi}, {Lawrence}, {Le Jeune}, {Levrier}, {Lilje},
  {Lilley}, {Lindholm}, {L{\'o}pez-Caniego}, {Ma}, {Mac{\'\i}as-P{\'e}rez},
  {Maggio}, {Maino}, {Mandolesi}, {Mangilli}, {Maris}, {Martin},
  {Mart{\'\i}nez-Gonz{\'a}lez}, {Matarrese}, {Mauri}, {McEwen}, {Melchiorri},
  {Mennella}, {Migliaccio}, {Miville-Desch{\^e}nes}, {Molinari}, {Moneti},
  {Montier}, {Morgante}, {Moss}, {Natoli}, {Oxborrow}, {Pagano}, {Paoletti},
  {Patanchon}, {Perdereau}, {Perotto}, {Pettorino}, {Piacentini},
  {Plaszczynski}, {Polastri}, {Polenta}, {Puget}, {Rachen}, {Racine},
  {Reinecke}, {Remazeilles}, {Renzi}, {Rocha}, {Rosset}, {Rossetti}, {Roudier},
  {Rubi{\~n}o-Mart{\'\i}n}, {Ruiz-Granados}, {Salvati}, {Sandri}, {Savelainen},
  {Scott}, {Sirignano}, {Sirri}, {Soler}, {Spencer}, {Suur-Uski}, {Tauber},
  {Tavagnacco}, {Tenti}, {Toffolatti}, {Tomasi}, {Tristram}, {Trombetti},
  {Valiviita}, {Van Tent}, {Vielva}, {Villa}, {Vittorio}, {Wandelt}, {Wehus},
  {Zacchei}, \& {Zonca}}]{Planck2016_GNILC}
{Planck Collaboration}, {Aghanim}, N., {Ashdown}, M., {et~al.}
  2016{\natexlab{a}}, \aap, 596, A109, \dodoi{10.1051/0004-6361/201629022}

\bibitem[{{Planck Collaboration} {et~al.}(2016{\natexlab{b}}){Planck
  Collaboration}, {Adam}, {Ade}, {Aghanim}, {Alves}, {Arnaud}, {Ashdown},
  {Aumont}, {Baccigalupi}, {Banday}, {Barreiro}, {Bartlett}, {Bartolo},
  {Battaner}, {Benabed}, {Beno{\^\i}t}, {Benoit-L{\'e}vy}, {Bernard},
  {Bersanelli}, {Bielewicz}, {Bock}, {Bonaldi}, {Bonavera}, {Bond}, {Borrill},
  {Bouchet}, {Boulanger}, {Bucher}, {Burigana}, {Butler}, {Calabrese},
  {Cardoso}, {Catalano}, {Challinor}, {Chamballu}, {Chary}, {Chiang},
  {Christensen}, {Clements}, {Colombi}, {Colombo}, {Combet}, {Couchot},
  {Coulais}, {Crill}, {Curto}, {Cuttaia}, {Danese}, {Davies}, {Davis}, {de
  Bernardis}, {de Rosa}, {de Zotti}, {Delabrouille}, {D{\'e}sert}, {Dickinson},
  {Diego}, {Dole}, {Donzelli}, {Dor{\'e}}, {Douspis}, {Ducout}, {Dupac},
  {Efstathiou}, {Elsner}, {En{\ss}lin}, {Eriksen}, {Falgarone}, {Fergusson},
  {Finelli}, {Forni}, {Frailis}, {Fraisse}, {Franceschi}, {Frejsel},
  {Galeotta}, {Galli}, {Ganga}, {Ghosh}, {Giard}, {Giraud-H{\'e}raud},
  {Gjerl{\o}w}, {Gonz{\'a}lez-Nuevo}, {G{\'o}rski}, {Gratton}, {Gregorio},
  {Gruppuso}, {Gudmundsson}, {Hansen}, {Hanson}, {Harrison}, {Helou},
  {Henrot-Versill{\'e}}, {Hern{\'a}ndez-Monteagudo}, {Herranz}, {Hildebrandt},
  {Hivon}, {Hobson}, {Holmes}, {Hornstrup}, {Hovest}, {Huffenberger}, {Hurier},
  {Jaffe}, {Jaffe}, {Jones}, {Juvela}, {Keih{\"a}nen}, {Keskitalo}, {Kisner},
  {Kneissl}, {Knoche}, {Kunz}, {Kurki-Suonio}, {Lagache},
  {L{\"a}hteenm{\"a}ki}, {Lamarre}, {Lasenby}, {Lattanzi}, {Lawrence}, {Le
  Jeune}, {Leahy}, {Leonardi}, {Lesgourgues}, {Levrier}, {Liguori}, {Lilje},
  {Linden-V{\o}rnle}, {L{\'o}pez-Caniego}, {Lubin}, {Mac{\'\i}as-P{\'e}rez},
  {Maggio}, {Maino}, {Mandolesi}, {Mangilli}, {Maris}, {Marshall}, {Martin},
  {Mart{\'\i}nez-Gonz{\'a}lez}, {Masi}, {Matarrese}, {McGehee}, {Meinhold},
  {Melchiorri}, {Mendes}, {Mennella}, {Migliaccio}, {Mitra},
  {Miville-Desch{\^e}nes}, {Moneti}, {Montier}, {Morgante}, {Mortlock}, {Moss},
  {Munshi}, {Murphy}, {Naselsky}, {Nati}, {Natoli}, {Netterfield},
  {N{\o}rgaard-Nielsen}, {Noviello}, {Novikov}, {Novikov}, {Orlando},
  {Oxborrow}, {Paci}, {Pagano}, {Pajot}, {Paladini}, {Paoletti}, {Partridge},
  {Pasian}, {Patanchon}, {Pearson}, {Perdereau}, {Perotto}, {Perrotta},
  {Pettorino}, {Piacentini}, {Piat}, {Pierpaoli}, {Pietrobon}, {Plaszczynski},
  {Pointecouteau}, {Polenta}, {Pratt}, {Pr{\'e}zeau}, {Prunet}, {Puget},
  {Rachen}, {Reach}, {Rebolo}, {Reinecke}, {Remazeilles}, {Renault}, {Renzi},
  {Ristorcelli}, {Rocha}, {Rosset}, {Rossetti}, {Roudier},
  {Rubi{\~n}o-Mart{\'\i}n}, {Rusholme}, {Sandri}, {Santos}, {Savelainen},
  {Savini}, {Scott}, {Seiffert}, {Shellard}, {Spencer}, {Stolyarov}, {Stompor},
  {Strong}, {Sudiwala}, {Sunyaev}, {Sutton}, {Suur-Uski}, {Sygnet}, {Tauber},
  {Terenzi}, {Toffolatti}, {Tomasi}, {Tristram}, {Tucci}, {Tuovinen}, {Umana},
  {Valenziano}, {Valiviita}, {Van Tent}, {Vielva}, {Villa}, {Wade}, {Wandelt},
  {Wehus}, {Wilkinson}, {Yvon}, {Zacchei}, \& {Zonca}}]{Planck2016Results}
{Planck Collaboration}, {Adam}, R., {Ade}, P.~A.~R., {et~al.}
  2016{\natexlab{b}}, \aap, 594, A10, \dodoi{10.1051/0004-6361/201525967}

\bibitem[{{Reach} \& {Heiles}(2021)}]{ReachHeiles2021}
{Reach}, W.~T., \& {Heiles}, C. 2021, \apj, 909, 71,
  \dodoi{10.3847/1538-4357/abd9c5}

\bibitem[{{Rezaei Kh.} \& {Kainulainen}(2022)}]{Rezaei2022}
{Rezaei Kh.}, S., \& {Kainulainen}, J. 2022, \apjl, 930, L22,
  \dodoi{10.3847/2041-8213/ac67db}

\bibitem[{{Rivilla} {et~al.}(2018){Rivilla}, {Jim{\'e}nez-Serra}, {Zeng},
  {Mart{\'\i}n}, {Mart{\'\i}n-Pintado}, {Armijos-Abenda{\~n}o}, {Viti},
  {Aladro}, {Riquelme}, {Requena-Torres}, {Qu{\'e}nard}, {Fontani}, \&
  {Beltr{\'a}n}}]{Rivilla2018}
{Rivilla}, V.~M., {Jim{\'e}nez-Serra}, I., {Zeng}, S., {et~al.} 2018, \mnras,
  475, L30, \dodoi{10.1093/mnrasl/slx208}

\bibitem[{{Rybarczyk} {et~al.}(2022{\natexlab{a}}){Rybarczyk}, {Gong},
  {Stanimirovi{\'c}}, {Babler}, {Murray}, {Winters}, {Luo}, {Dame}, \&
  {Steffes}}]{Ryb2_2022}
{Rybarczyk}, D.~R., {Gong}, M., {Stanimirovi{\'c}}, S., {et~al.}
  2022{\natexlab{a}}, \apj, 926, 190, \dodoi{10.3847/1538-4357/ac4160}

\bibitem[{{Rybarczyk} {et~al.}(2022{\natexlab{b}}){Rybarczyk},
  {Stanimirovi{\'c}}, {Gong}, {Babler}, {Murray}, {Gerin}, {Winters}, {Luo},
  {Dame}, \& {Steffes}}]{Ryb1_2022}
{Rybarczyk}, D.~R., {Stanimirovi{\'c}}, S., {Gong}, M., {et~al.}
  2022{\natexlab{b}}, \apj, 928, 79, \dodoi{10.3847/1538-4357/ac5035}

\bibitem[{{Schilke} {et~al.}(2001){Schilke}, {Pineau des For{\^e}ts},
  {Walmsley}, \& {Mart{\'\i}n-Pintado}}]{Schilke2001}
{Schilke}, P., {Pineau des For{\^e}ts}, G., {Walmsley}, C.~M., \&
  {Mart{\'\i}n-Pintado}, J. 2001, \aap, 372, 291,
  \dodoi{10.1051/0004-6361:20010470}

\bibitem[{{Schilke} {et~al.}(1997){Schilke}, {Walmsley}, {Pineau des Forets},
  \& {Flower}}]{Schilke1997}
{Schilke}, P., {Walmsley}, C.~M., {Pineau des Forets}, G., \& {Flower}, D.~R.
  1997, \aap, 321, 293

\bibitem[{{Sch{\"o}ier} {et~al.}(2010){Sch{\"o}ier}, {van der Tak}, {van
  Dishoeck}, \& {Black}}]{2010ascl.soft10077S}
{Sch{\"o}ier}, F., {van der Tak}, F., {van Dishoeck}, E., \& {Black}, J. 2010,
  {LAMDA: Leiden Atomic and Molecular Database}.
\newblock \doeprint{1010.077}

\bibitem[{{Snow} \& {McCall}(2006)}]{SnowMcCall2006}
{Snow}, T.~P., \& {McCall}, B.~J. 2006, \araa, 44, 367,
  \dodoi{10.1146/annurev.astro.43.072103.150624}

\bibitem[{{Spezzano} {et~al.}(2020){Spezzano}, {Codella}, {Podio},
  {Ceccarelli}, {Caselli}, {Neri}, \& {L{\'o}pez-Sepulcre}}]{Spezzano2020}
{Spezzano}, S., {Codella}, C., {Podio}, L., {et~al.} 2020, \aap, 640, A74,
  \dodoi{10.1051/0004-6361/202037864}

\bibitem[{{Turner}(1998)}]{Turner1998}
{Turner}, B.~E. 1998, \apj, 495, 804, \dodoi{10.1086/305319}

\bibitem[{{van der Tak} {et~al.}(2007){van der Tak}, {Black}, {Sch{\"o}ier},
  {Jansen}, \& {van Dishoeck}}]{vanderTak2007}
{van der Tak}, F.~F.~S., {Black}, J.~H., {Sch{\"o}ier}, F.~L., {Jansen}, D.~J.,
  \& {van Dishoeck}, E.~F. 2007, \aap, 468, 627,
  \dodoi{10.1051/0004-6361:20066820}

\bibitem[{{Wiklind} \& {Combes}(1997)}]{WilkindCombes1997}
{Wiklind}, T., \& {Combes}, F. 1997, \aap, 324, 51,
  \dodoi{10.48550/arXiv.astro-ph/9701081}

\bibitem[{{Yan} {et~al.}(2004){Yan}, {Lazarian}, \& {Draine}}]{Yan2004}
{Yan}, H., {Lazarian}, A., \& {Draine}, B.~T. 2004, \apj, 616, 895,
  \dodoi{10.1086/425111}

\bibitem[{{Yu} {et~al.}(2018){Yu}, {Xu}, \& {Wang}}]{NaiPing2018}
{Yu}, N.-P., {Xu}, J.-L., \& {Wang}, J.-J. 2018, Research in Astronomy and
  Astrophysics, 18, 015, \dodoi{10.1088/1674-4527/18/2/15}

\bibitem[{{Zhu} {et~al.}(2017){Zhu}, {Tian}, {Li}, \& {Zhang}}]{Zhu2017}
{Zhu}, H., {Tian}, W., {Li}, A., \& {Zhang}, M. 2017, \mnras, 471, 3494,
  \dodoi{10.1093/mnras/stx1580}

\end{thebibliography}
\bibliographystyle{aasjournal}

\end{document}